\documentclass[12pt]{iopart}
\usepackage{iopams,epsfig}
\usepackage{booktabs}
\usepackage{units,cite}

\renewcommand{\th}{\mathrm{th}}

\begin{document}

\title[On inconsistency of experimental data on primary nuclei]{On inconsistency
of experimental data on primary
nuclei spectra with sea level muon intensity measurements}

\author{A A Lagutin, A G Tyumentsev and A V Yushkov}
\address{Theoretical Physics Department, Altai State University,
Lenin Avenue 61, Barnaul 656049, Russia}
\ead{yushkov@theory.dcn-asu.ru}

\begin{abstract}
For the first time a complete set of the most recent direct data on
primary cosmic ray spectra is used as input into calculations of muon
flux at sea level in wide energy range $E_\mu=1-3\cdot10^5$~GeV.
Computations have been performed with the CORSIKA/QGSJET and
CORSIKA/VENUS codes. The comparison of the obtained muon intensity
with the data of muon experiments shows, that measurements of primary
nuclei spectra conform to sea level muon data only up to several tens
of GeV and result in essential deficit of muons at higher energies.
As it follows from our examination, uncertainties in muon flux
measurements and in the description of nuclear cascades development
are not suitable to explain this contradiction, and the only
remaining factor, leading to this situation, is underestimation of
primary light nuclei fluxes. We have considered systematic effects,
that may distort the results of the primary cosmic ray measurements
with the application of the emulsion chambers. We suggest, that
re-examination of these measurements is required with the employment
of different hadronic interaction models. Also, in our point of view,
it is necessary to perform estimates of possible influence of the
fact, that sizable fraction of events, identified as protons,
actually are antiprotons. Study of these cosmic ray component begins
to attract much attention, but today nothing definite is known for
the energies $>40$~GeV. In any case, to realize whether the
mentioned, or some other reasons are the sources of disagreement of
the data on primaries with the data on muons, the indicated effects
should be thoroughly analyzed.
\end{abstract}


\vfill

{\sl Accepted for publication in \JPG}

\maketitle

\section{Introduction}
A muon component of cosmic rays plays an important role in many
fields of astroparticle physics. It provides a basis for verification
of our knowledge on primary cosmic ray (PCR) spectrum behaviour,
high-energy hadronic interactions and for solution of neutrino
physics problems. In view of this, a question of interconsistency of
information, gained in the last two decades on primary and muon
fluxes, and on high-energy hadronic interactions presents great
importance. In well known works of Volkova et al. (1979)~\cite{volk},
Dar (1983)~\cite{dar}, Butkevich et al.~(1989)~\cite{butkevich},
Lipari (1993)~\cite{lipari}, Honda et al. (1995)~\cite{honda},
Agrawal et al. (1996)~\cite{agrawal} and Bugaev et al.
(1998)~\cite{bugaev} this problem was outside of the consideration,
since the principal aim lied in the estimation of the secondary
lepton fluxes. Authors relied on the information on primaries and
nuclear interactions, that was available at the papers writing time.
Certainly, these data were incomplete and ambiguous, as the
consequence the inputs in the calculations also vary rather
significantly. Nevertheless, the outputs, i.e. muon fluxes,
satisfactory agree with each other (see comparisons
in~\cite{honda,agrawal,bugaev}) and muon experimental data. The given
circumstance with necessity shows, that discrepancies in the used PCR
spectra in the large part were compensated by different approaches to
the treatment of nuclear cascades. This, in turn, may possibly relate
to the fact, that muon flux in most of these works served only for
normalization of neutrino fluxes and it was fitted to the muon
experimental data via adjustment of not precisely known parameters of
hadronic interactions. From these short remarks it is clear, that all
these calculations lack some standardization, as it was recently
proposed by Gaisser and Honda~\cite{gaisser2002} in respect to the
PCR spectra.

The current situation provides significantly better capabilities for
adequate choice of PCR model and for simulation of nuclear cascades,
and in this paper we have tried to take the maximum advantage of
this. But, as our analysis has shown, unfortunately the information
on all components of calculations is still very uncertain to provide
a firm ground for accurate derivation of muon flux at sea level and,
today, it is following `from top to bottom'. Direct measurements of
primary cosmic ray spectra span up to the energy of $\sim$1~PeV for
protons and up to few hundred TeV/n for other groups of nuclei. Most
abundant data are collected for $E_\mathrm{PCR}<1$~TeV/n. Here proton
spectrum is studied in series of recent satellite and balloon
experiments with approximately 20\% accuracy. Higher energy data of
SOKOL~\cite{sokol}, MUBEE~\cite{zatsepin}, JACEE~\cite{jacee} and
RUNJOB~\cite{runjob} although have relatively large statistic and
systematic errors, also satisfactory agree with each other. At the
same time, the measurements of the helium spectrum are much less
concordant and still differ by almost a factor of 2 for all energies,
and this uncertainty is especially crucial for evaluation of
$\mu^+/\mu^-$ and $\tilde\nu/\nu$ ratios. As it is widely known,
protons and helium nuclei contribute $\sim90\%$ to the nucleon flux
at the top of the atmosphere, which is relevant for derivation of the
muon spectrum at sea level, thus the particular emphasis should be
given to the accurate description of these spectra.

To avoid simplification, for the simulation of nuclear cascades we
have applied widely approved and thoroughly tested Monte-Carlo code
CORSIKA~\cite{corsika}, that allows to treat hadronic interactions
with the use of any of the up-to-date interaction models:
QGSJET~\cite{qgsjet}, VENUS~\cite{venus}, HDPM~\cite{corsika},
NeXuS~\cite{nexus} or DPMJET~\cite{dpmjet}. Overall uncertainty,
brought in computations of muon flux by the use of these models, is
not very significant and considerably decreased in the last decade.
So, the difference in $p-air$ inelastic cross-sections `between the
models have shrunk from 80~mb to today 20~mb in the region of few
PeV'~\cite{heck2002}. Average numbers of muons in cascades from
primary protons, obtained with the use of these models, differ not
more than by 20\%~\cite{fzka5828} for all energies of interest. This
means, that integral and differential muon fluxes at sea level will
also differ by a close value, and this discrepancy is the smallest,
in comparison with the uncertainties of PCR chemical composition and
energy spectra for $E_\mathrm{PCR}\gtrsim 1$~TeV/n, and the sea level
muon data. The latter, for $E_\mu\lesssim100$~GeV, are both numerous
and, on the whole, quite ambiguous (see extensive summary
in~\cite{summary}), but the most recent ones of BESS~\cite{bess_mu}
and CAPRICE~\cite{caprice_mu} are very precise and closely agree with
each other. They give a basis for checking information on primary
light nuclei spectra up to the energies $\sim10^4$~GeV/n. The
interval of higher muon energies $10^2-10^4$~GeV, where the spread of
the data on muon intensity is still small and amounts to some 20\%,
allows to examine with nearly the same accuracy behaviour of primary
proton and helium spectra for the energies, extending to the upper
bound of balloon and satellite measurements $\sim1$~PeV/n. Evidently,
that beyond the PCR data, it is not feasible to perform reliable
calculations of the muon spectrum. What is more, the inverse problem,
i.e. reconstruction of PCR fluxes from the data on muon spectrum, can
not be also solved, because for $E_\mu>10$~TeV there are only
indirect data, in which muon flux at sea level is derived from
results of underground measurements. These data are quite
contradictory and have large systematic errors, mainly caused by
incomplete information on rock properties and vagueness in question
of prompt muon generation mechanisms (see, for example,
\cite{markov,bugaev}). As a consequence of all this, now there is no
possibility to make conclusions neither on the preferability of any
model of charm generation, nor on behaviour of PCR spectra for
$E_\mathrm{PCR}>1$~PeV/n.

As it is stated above, this paper is devoted to the investigation of
compatibility of the present sea level muon flux measurements in
energy interval $E_\mu=1-10^5$~GeV with the experimental data on PCR
spectra for corresponding primary energies $10-10^6$~GeV/n. In
section~2 we briefly review the present data on primaries of H, He,
CNO, Ne-Si and Fe groups and discuss PCR models, applied in this
paper and in the papers of the other authors. Basic characteristics
of computations are presented in section~3. Section~4 is devoted to
consideration of consistency of our vertical muon flux, derived from
the data of direct PCR measurements, with sea level muon experiments
in the energy range $E_\mu=1-10^5$~GeV. Since in this section we have
revealed a sizable shortage of muons with energies, corresponding to
that of primaries, studied with the emulsion chambers, in section~5
we examine possible effects, that may lead to a distortion of the
information, obtained in the space and balloon experiments, employing
this  technique. Our conclusions are given in section~6.

\section{Primary cosmic ray spectra}

\begin{figure}[h]
\centering\includegraphics[height=.4\textheight]{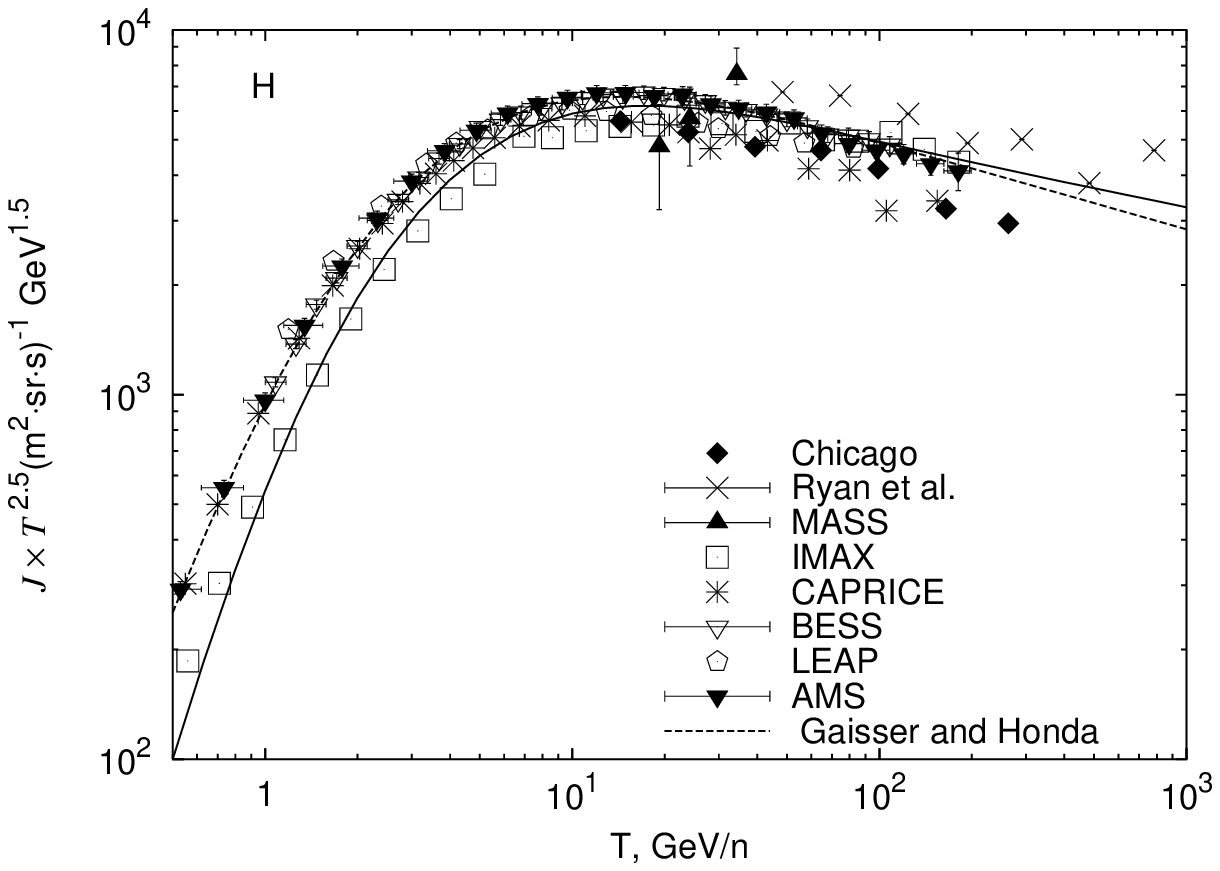}\hfill
\centering\includegraphics[height=.4\textheight]{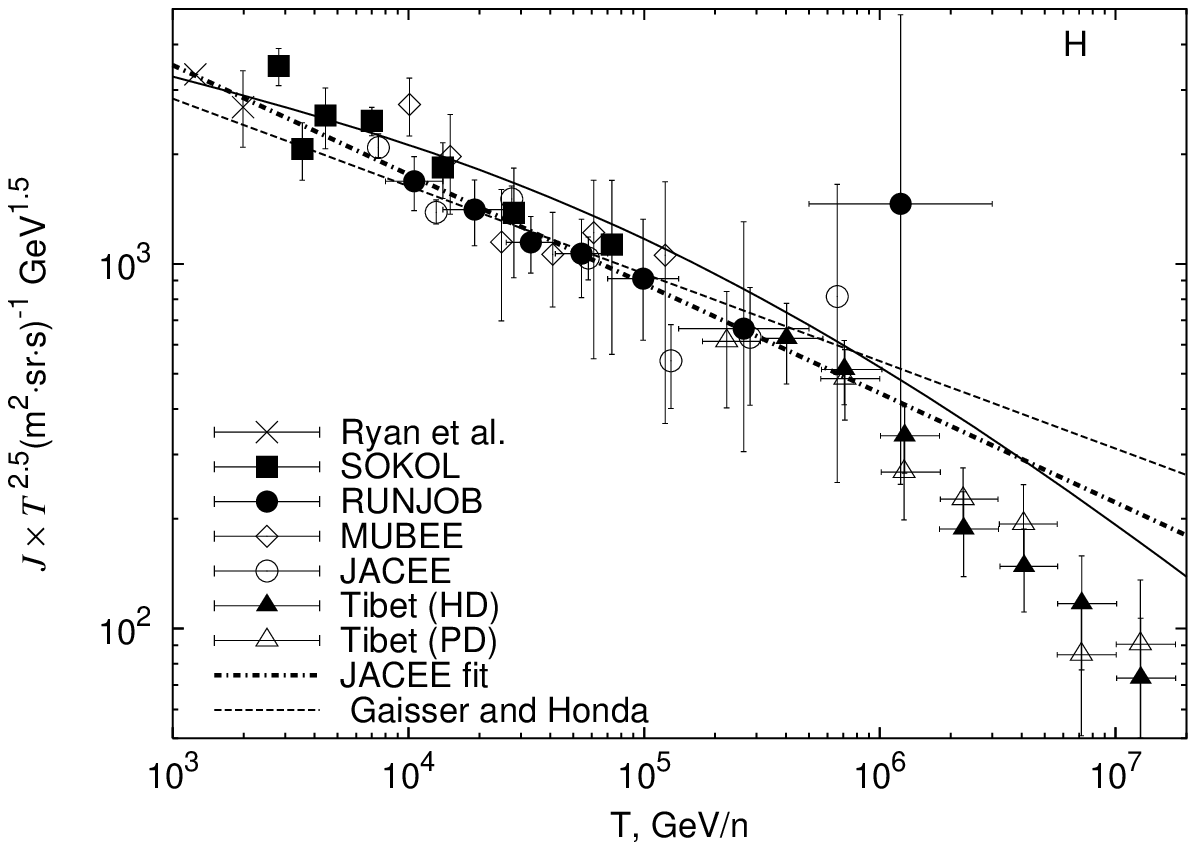}
\caption{Primary proton differential spectrum. Experimental data:
\cite{swordy}~Chicago, \cite{ryan}~Ryan et al., \cite{papini}~MASS,
\cite{sokol}~SOKOL, \cite{imax}~IMAX, \cite{caprice}~CAPRICE,
\cite{runjob}~RUNJOB, \cite{bess}~BESS, \cite{jacee}~JACEE and JACEE
fit, \cite{leap}~LEAP, \cite{zatsepin}~MUBEE, \cite{ams}~AMS,
\cite{tibet2003}~Tibet~(HD) and Tibet~(PD). Dashed line is the
spectrum, proposed by Gaisser and Honda~\cite{gaisser2002}. Solid
line is the spectrum, used in this paper.} \label{p}
\end{figure}

\begin{figure}[h]
\centering\includegraphics[height=.4\textheight]{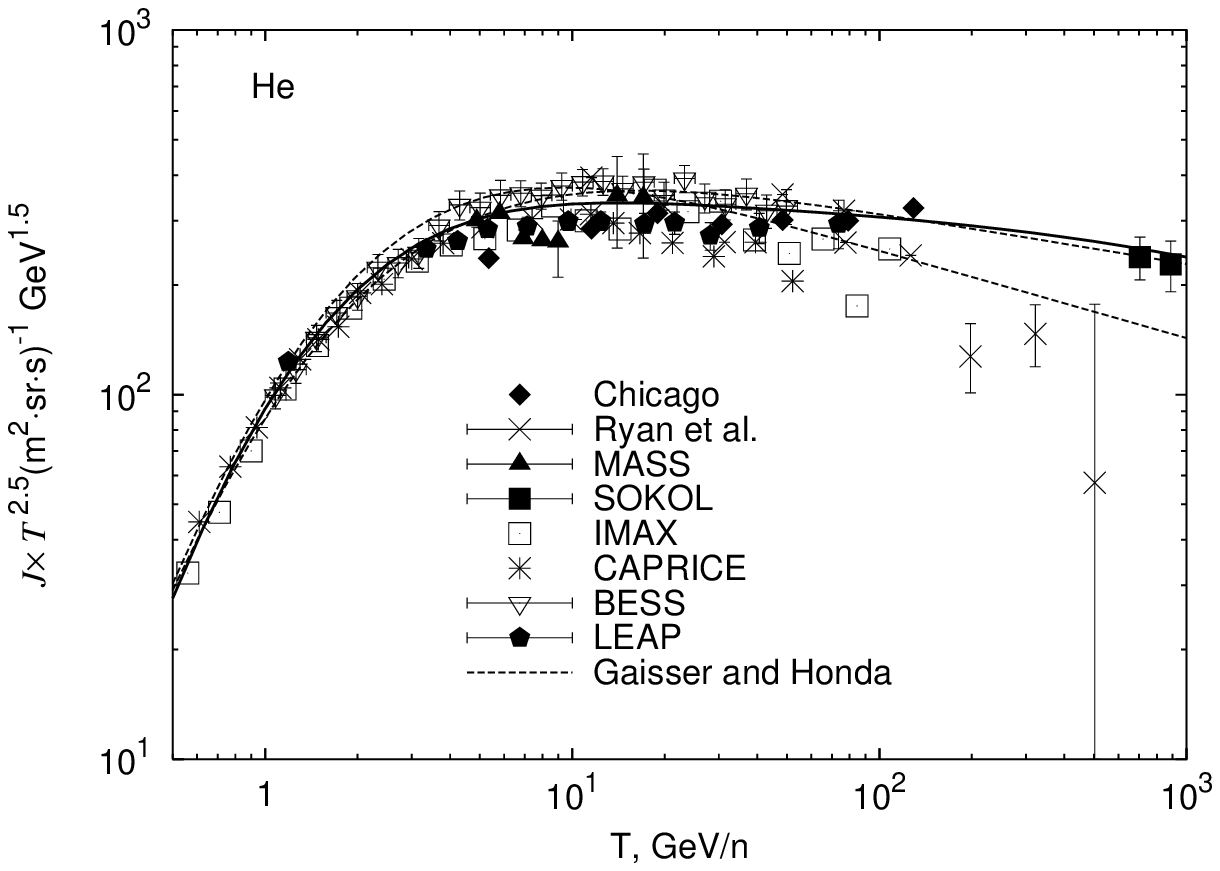}\hfill
\centering\includegraphics[height=.4\textheight]{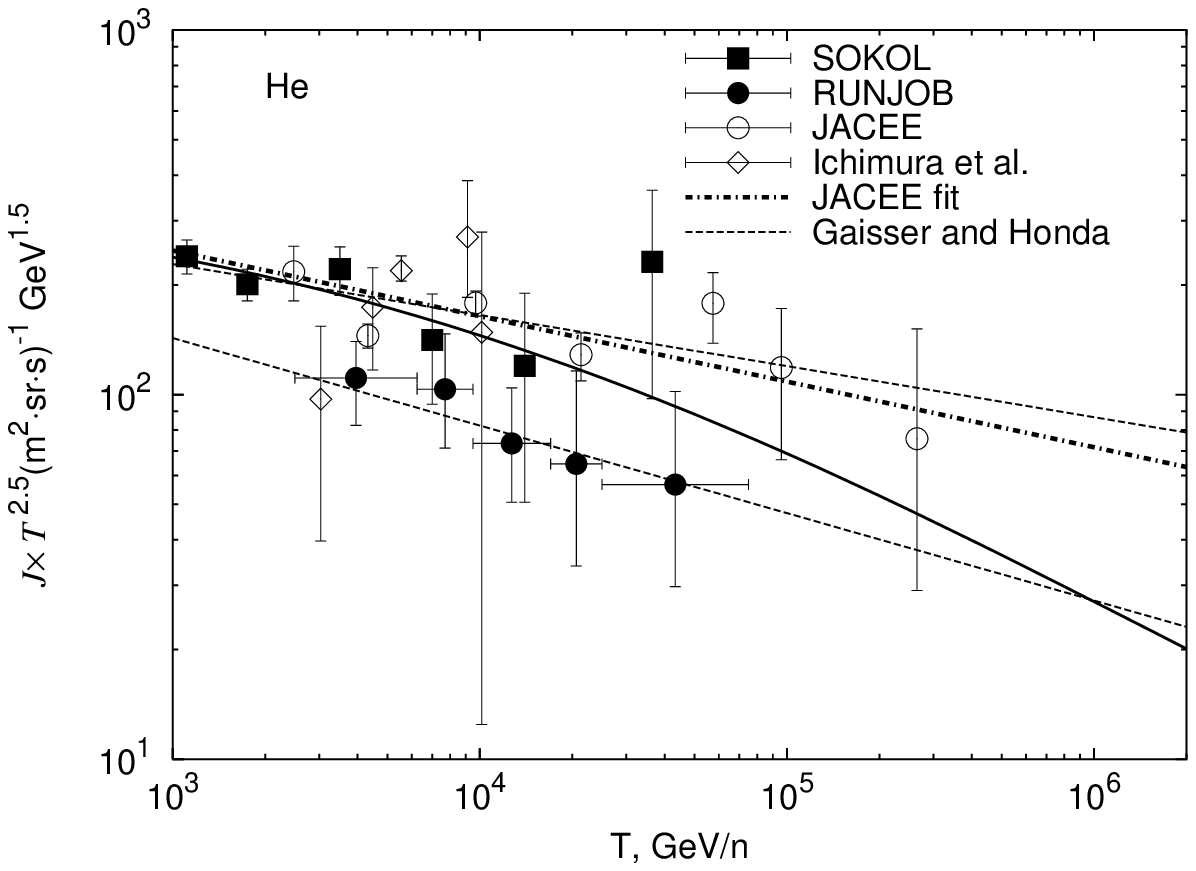}
\caption{Primary helium differential spectrum. Experimental data:
\cite{ichimura}~Ichimura et al. Two lines for the Gaisser and Honda
spectrum correspond to their `low' and `high' helium fits. Other
designations are the same as in figure~\ref{p}.} \label{he}
\end{figure}

\begin{figure}
\centering\includegraphics[height=.3\textheight]{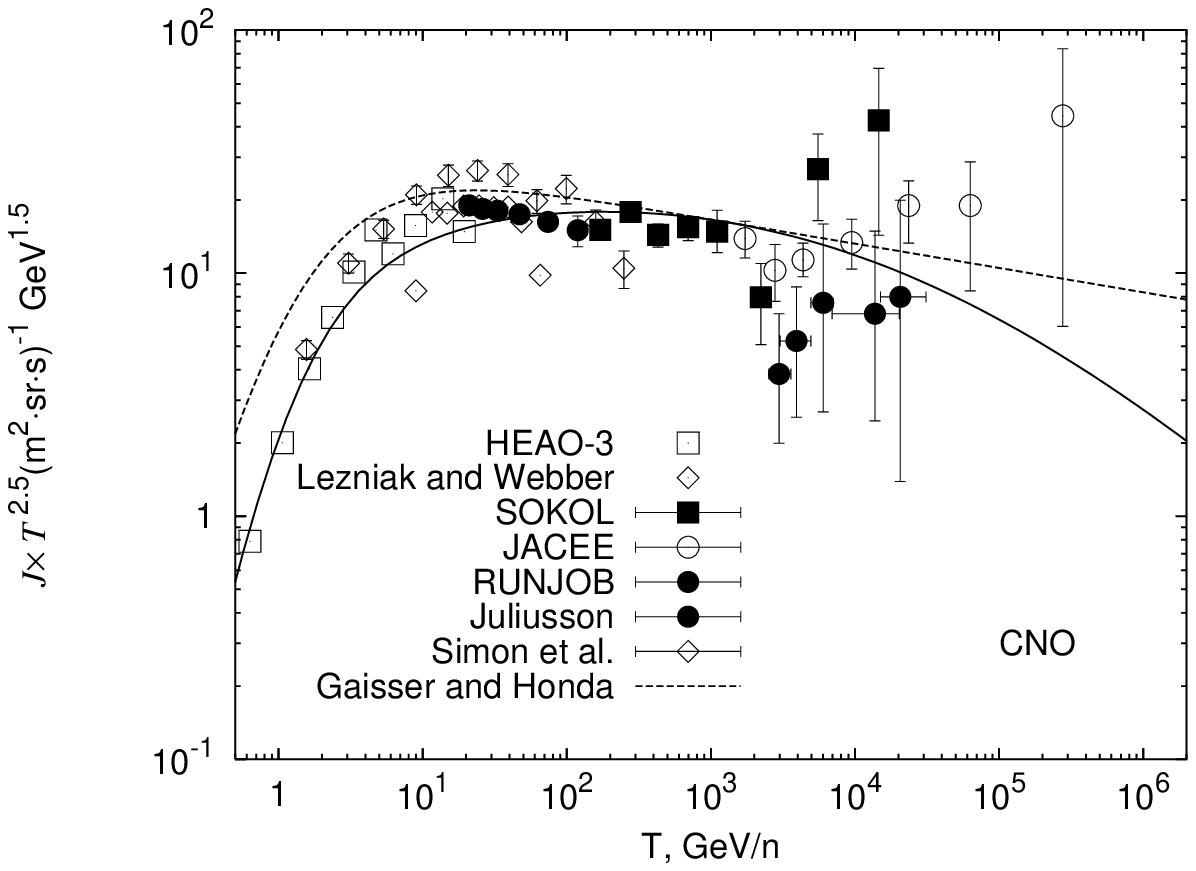}\hfill
\centering\includegraphics[height=.3\textheight]{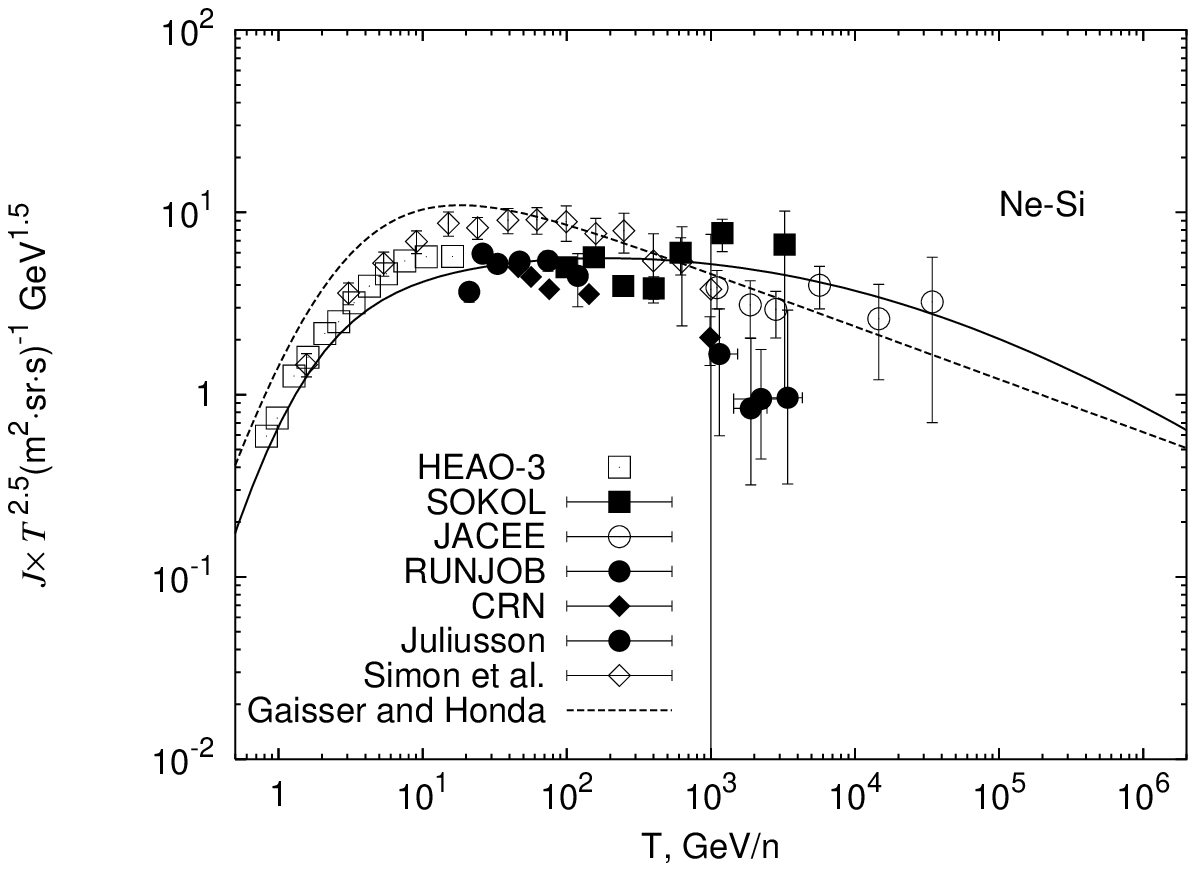}
\centering\includegraphics[height=.3\textheight]{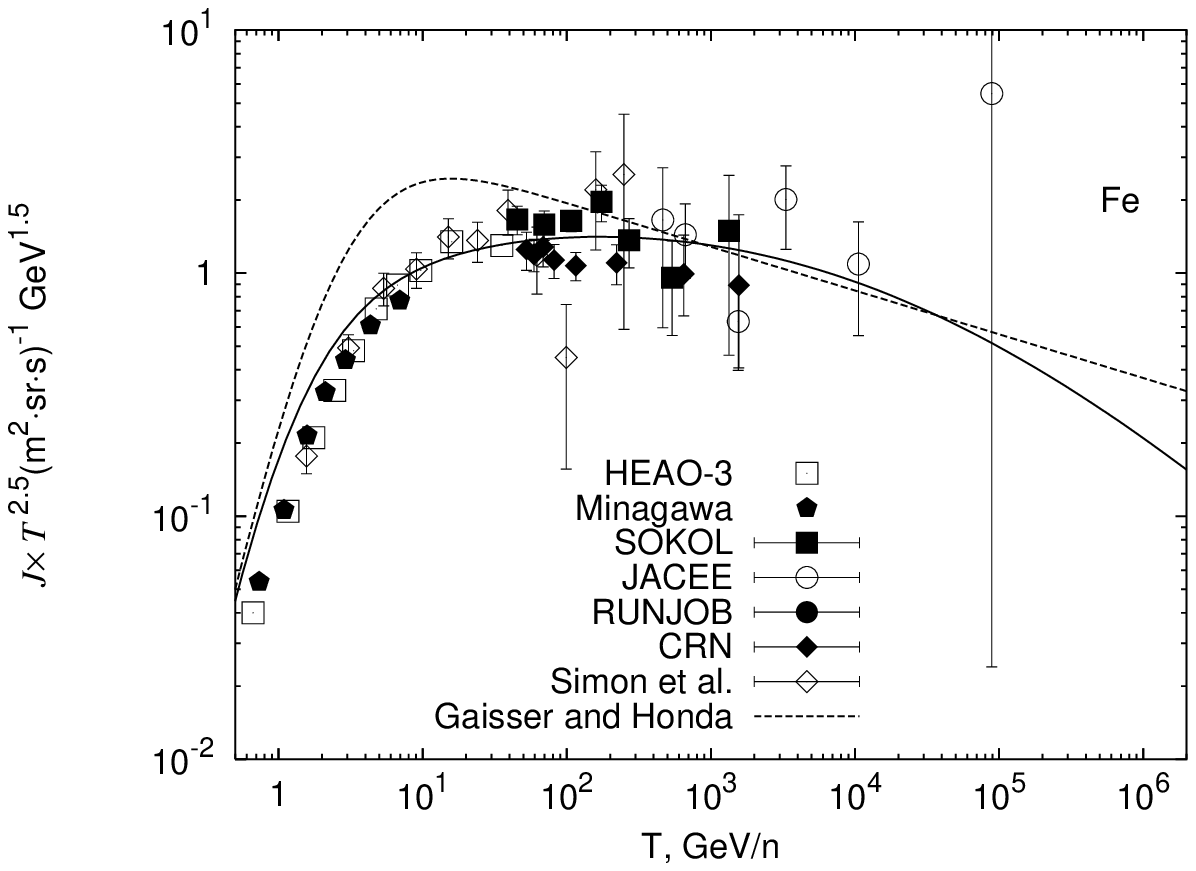}\hfill\caption{Primary
nuclei differential spectra. Experimental data: \cite{heao}~HEAO-3,
\cite{lezniak}~Lezniak and Webber, \cite{juliusson}~Juliusson,
\cite{simon}~Simon et al., \cite{crn}~CRN, \cite{minagawa}~Minagawa.
Other designations are the same as in figure~\ref{p}.} \label{hefe}
\end{figure}

The extensive compilation of the modern data of space and balloon
measurements plotted vs kinetic energy $T,$~GeV/n is shown in
figures~\ref{p}-\ref{hefe}. The most of the data are gathered for
relatively low energies $<1$~TeV/n. But even here some 15 years ago
the spread in experimental results was at least 100\% for all groups
of nuclei. In spite of series of experiments, performed since then,
significant improvement was achieved only in the study of the proton
spectrum. So, for energies from $\sim$10 GeV, where effects of
geomagnetic cutoff and solar modulations become negligible, to some
hundred GeV, data of recent experiments LEAP~\cite{leap},
CAPRICE~\cite{caprice}, IMAX~\cite{imax}, BESS~\cite{bess} and
AMS~\cite{ams} are consistent within 20\% (see figure~\ref{p}).
Better mutually consistent, within 5--10\%, data of AMS and BESS, now
are considered as the most precise. From $\sim$~200~GeV to
$\sim2$~TeV there is a gap between magnet spectrometer and emulsion
chamber experimental data, with the only ionization calorimeter
measurements of Ryan et al.~\cite{ryan}. Since at the energies,
overlapping with AMS and BESS, the given data overestimate proton
intensity (but not for helium), probably they should be lowered by
25\%, as proposed in~\cite{gaisser2001}. In any case, more accurate
information on behaviour of the proton spectrum in this energy
interval is required for reliable evaluation of the muon and neutrino
fluxes for the energies below 300~GeV. From approximately 2~TeV to
$100$~TeV SOKOL, MUBEE, JACEE and RUNJOB measurements provide rather
consistent information on proton flux, though with the nearly 20\%
uncertainty. For understanding of the proton spectrum behaviour
beyond the `knee', valuable information is provided by the pioneer
measurements of the Tibet hybrid experiment~\cite{tibet2003}, carried
out at the mountain altitude. These first results point at the
steepening of the proton spectrum for $E_p\gtrsim200$~TeV and this is
an essential guideline for the extrapolation of the direct
measurements. Spread in the data for all heavier nuclei practically
for all energies of interest is much larger than that for hydrogen,
and equals to 100\% and more. The difference between the helium
spectra of SOKOL, JACEE and RUNJOB is of major importance for
accurate evaluation of secondary lepton fluxes.

\begin{figure}
\centering\includegraphics[height=.3\textheight]{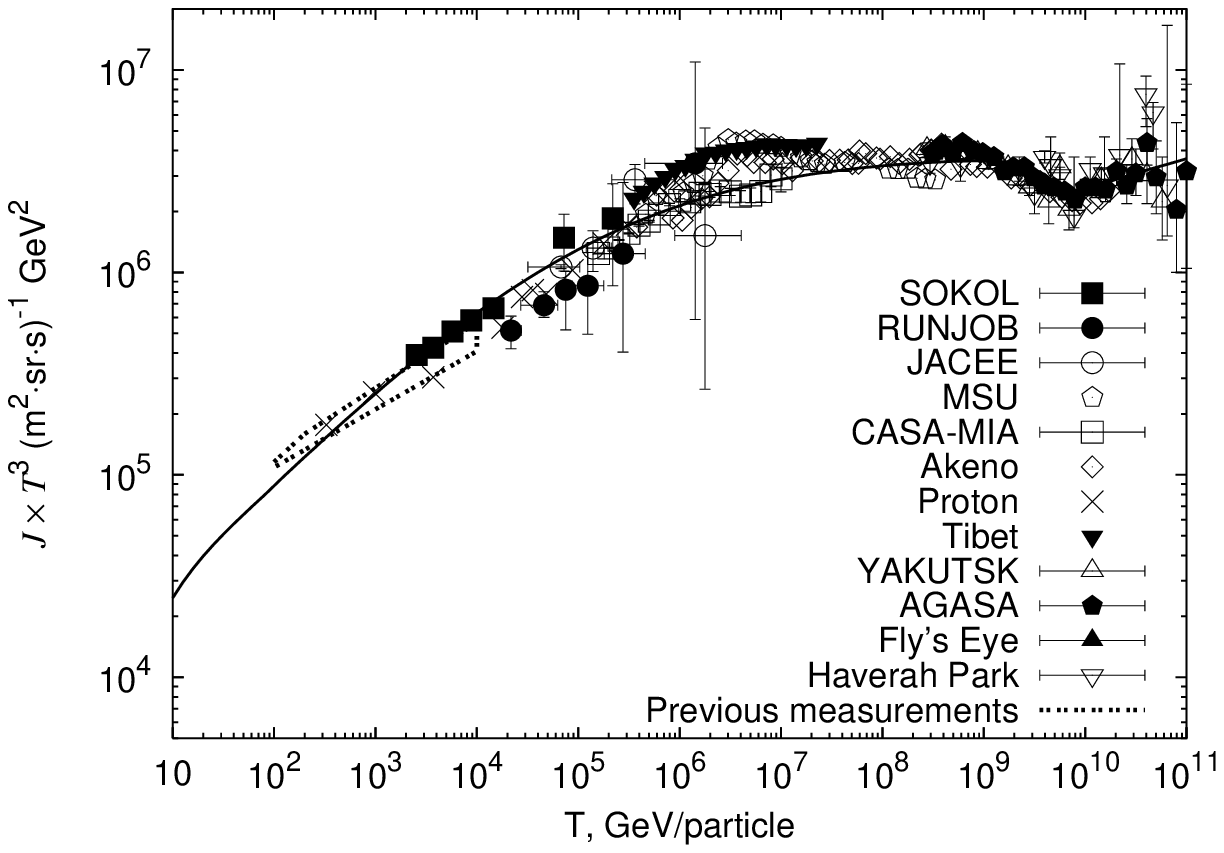}
\caption{All-particle primary spectrum. Experimental data:
\cite{sokol}~SOKOL, \cite{runjob}~RUNJOB, \cite{jacee}~JACEE,
\cite{msu}~MSU, \cite{casa-mia}~CASA-MIA, \cite{akeno}~AKENO,
\cite{grigorov}~Proton, \cite{tibet96}~Tibet. YAKUTSK, AGASA, Fly's
Eye and Haverah Park are cited according to \cite{cronin}. `Previous
measurements' are taken from~\cite{linsley}. Solid line is the
spectrum, used in this paper.} \label{allp}
\end{figure}

From the aforesaid it becomes evident, that today no unique fit of
all data on primary spectra can be given and considerable
arbitrariness in the choice of them still remains. In such events one
may use, for example, upper and lower estimates of all primary
fluxes, or the flux, giving some average of the experimental data.
The principal model of the PCR spectrum, employed in our
calculations, may be regarded just as the latter case. It was
obtained in frameworks of anomalous diffusion of cosmic rays in
fractal interstellar medium~\cite{lagutin2001,bf,bf2003,lagutin2003},
but here it is considered irrespectively to its validity for
description of cosmic ray propagation. Altogether in our computations
we have accounted for 5 groups of primaries: H, He, CNO, Ne-Si and
Fe. Mass composition of the applied model is fitted to the direct and
EAS experimental data on the all-particle primary spectrum
(figure~\ref{allp}). As it it seen from figures~\ref{p}-\ref{hefe}
our model as well satisfactory corresponds to the available
experimental information on elemental spectra and for high energies
really gives some average of the data on nuclei, except that for
hydrogen it presents rather an upper estimate.  For calculation of
muon flux it also matters, that the data on the proton component
become insufficiently statistically provided from $\sim10^5$~GeV, and
that sizeable uncertainty gives $\pm50$\% spread of the data on
helium of JACEE and RUNJOB around the values, adopted in our model.
So, though on the whole the situation looks quite well-defined for
calculations of vertical muon intensity up to the energies of several
TeV, for the greater energies the existing PCR data are very
ambiguous. In order to estimate the influence of these data spread on
the resulting  muon flux, we have additionally performed calculations
with the use of another two PCR models, shown in
figures~\ref{p}-\ref{hefe}: first of them presents the JACEE fit of
their hydrogen and helium measurements~\cite{jacee}, and the second
is the model, suggested by Gaisser and Honda~\cite{gaisser2002}.

\begin{figure}
\centering\includegraphics[height=.4\textheight]{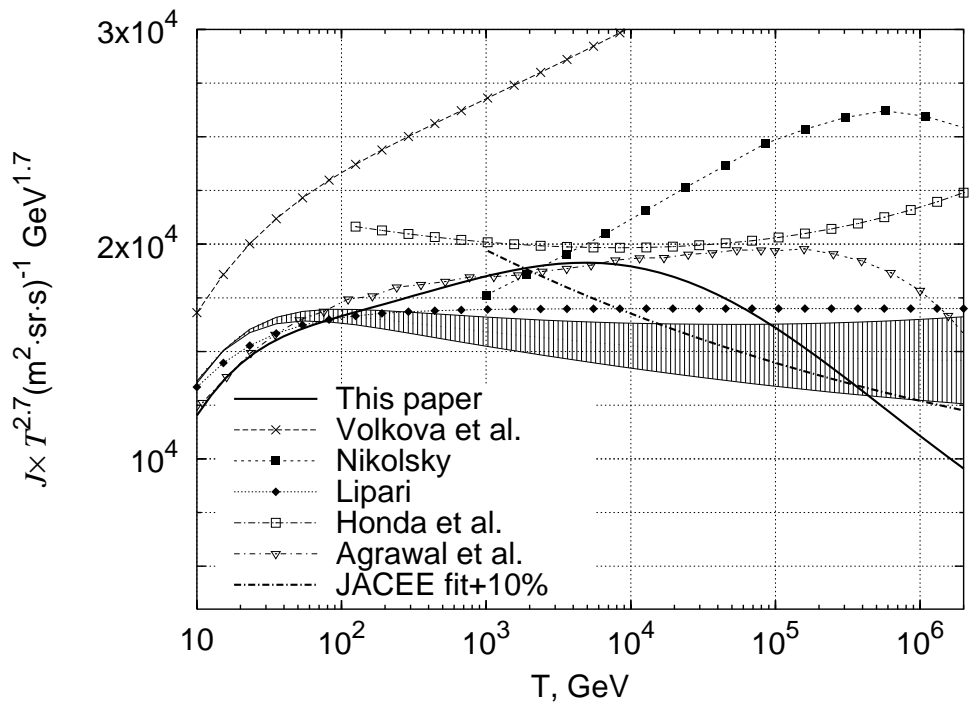}
\caption{Primary nucleon differential spectra. \cite{volk}~Volkova et
al.; \cite{NSU}~Nikolsky primary spectrum, applied for calculations
of muon flux in paper of Bugaev et al.~\cite{bugaev};
\cite{lipari}~Lipari; \cite{honda}~Honda et al.;
\cite{agrawal}~Agrawal et al.; \cite{jacee}~JACEE fit+10\%. Shaded
area is the spectrum, proposed by Gaisser and
Honda~\cite{gaisser2002}, with the lower and upper bounds
corresponding to the `high' and `low' helium fits. Solid line is the
spectrum, used in this paper.} \label{nucleons}
\end{figure}

To compute sea level muon flux we have converted primary nuclei
spectra to a spectrum in nucleons in standard
manner:
\[
J_N(E)=J_p(E)+\sum_A J_A(E\cdot A)\cdot A^2,
\]
here $J_A(E)$ is the differential energy spectrum of the nuclei with
atomic number $A$, $A=4,14,28,56$. In figure~\ref{nucleons} it is
presented, along with our primary nucleon flux (solid line), fluxes,
applied in calculations of muon intensity in
works~\cite{volk,lipari,honda,agrawal,bugaev}. These PCR spectra vary
both in shapes and values and here it is useful to discuss all of
them in order to realize, how they correspond to the modern
experimental data. First of all, one can see that a sizable
discrepancy between our, `Gaisser and Honda' and `JACEE fit + 10\%'
(here we add 10\% to account a contribution of heavier, than H and
He, nuclei) prevails, and this is a good illustration of the
difficulties in the attaining of the unique description of the
current experimental data. Behaviour of our model requires some more
comments. Enhancement by $\sim20\%$ of intensity of the proton
component in the energy range $10^3-10^5$~GeV is dictated by the
fact, that our first test calculations of muon flux with the
`average', close to JACEE fit, proton flux resulted in a very large
shortage of muons at sea level (see section~4). In order to smooth
this contradiction we have tested several variants of PCR spectra and
eventually adopted the presented version. It gives the maximal
increase of the nucleon flux, provided the elemental and all-particle
spectra stay consistent with the experiment. Note, that if the muon
data did not force us to enlarge the hydrogen flux, then it would
reasonably agree with the new data of the Tibet AS$\gamma$
collaboration~\cite{tibet2003}. Dip of our nucleon and elemental
fluxes for $E_\mathrm{PCR}\gtrsim10^5$~GeV/n is an intrinsic feature
of the anomalous diffusion propagation model, and, on the other hand,
it corresponds to the numerous indications about the change of the
PCR spectral index in the `knee' region.

From the earlier works, the most appropriate PCR models were employed
by Lipari~\cite{lipari} and Agrawal et~al.~\cite{agrawal}. The
nucleon spectrum from the latter paper almost coincides with our one,
deviating only from the energy of about $10^4$~GeV. Mainly, this
discrepancy is caused by the use of excessive, in comparison with
today's, flux of protons, obtained from the results of the first 8
JACEE flights~\cite{jacee1993}. In 1993 JACEE reported, that proton
spectrum consists of two parts with different slopes and a spectral
break at $E_p=40$~TeV. In paper~\cite{agrawal} this break was not
accounted for and extrapolation of the data for $E_p<40$~TeV was
applied for the higher energies. In work of Lipari a simple power
primary nucleon spectrum was picked, which is convenient to use in
analytic calculations, and no collation with the experiment was
presented. Nevertheless, this is a rather appropriate model, closely
agreeing with the `Gaisser and Honda' fit.

The three remaining spectra, applied by Volkova et al.~\cite{volk},
Honda et al.~\cite{honda} and Bugaev et al.~\cite{bugaev} are quite
excessive, compared to what is known today. In the oldest of the
discussed work of Volkova et al.~\cite{volk}, it is applied very
large all-nucleon spectrum $J_N(E)=1.9\cdot10^4E^{-2.65}$
$(\mathrm{GeV}\cdot\mathrm{m}^2\cdot\mathrm{sec}\cdot\mathrm{sr})^{-1}$,
obtained from EAS data, and no elemental composition had been
considered. Spectrum, used by Bugaev et al.~\cite{bugaev}, is taken
from the work of Nikolsky~\cite{NSU}, where it was derived from the
analysis of fluctuations of muon and electron numbers in EAS. For
$E_p>10^4$~GeV proton intensity in this model is essentially
overestimated (as a consequence, nucleon spectrum too), in comparison
with the experimental data, and precisely such behaviour provides a
good agreement of the resulting muon flux in~\cite{bugaev} with the
most of experimental data on sea level muon intensity. In the paper
of Honda et al.~\cite{honda} the proton flux for all energies  is
also rather high for the following reasons. In the low energy region
the given spectrum relies on the compilation of Webber and
Lezniak~\cite{webber}, i.e. mostly on the Ryan et al.~\cite{ryan}
data. But they are referred in~\cite{webber} incorrectly and do not
match the original values, overstating them by $\sim20\%$.
Extrapolation of this spectrum to the higher energies reasonably
agree with the JACEE~$1-8$ flights results, but, as it was already
said, to some extent overvalues the present data. From the analysis
of these three last models a simple deduction follows: their
renormalization down to the experimental values should cause a
sizable deficit of calculated sea level muons. We shall return to
this statement again in section~4.

\section{Basic characteristics of calculations}

Simulation of cascade processes in the atmosphere has been performed
with the use of CORSIKA (v6.00, v6.018). In most of calculations as
the model describing hadron-nucleus interaction for energies
$E_\mathrm{lab}>80$~GeV QGSJET model is applied. This model is in a
good agreement with accelerator and EAS experimental data (see
e.g.~\cite{fzka5828,agasa}) and provides comparatively high
calculation speed. Hadronic interactions with energies
$E_\mathrm{lab}<80$~GeV were simulated with GHEISHA~\cite{gheisha}.

\begin{figure}[h]
\centering\includegraphics[width=.49\textwidth]{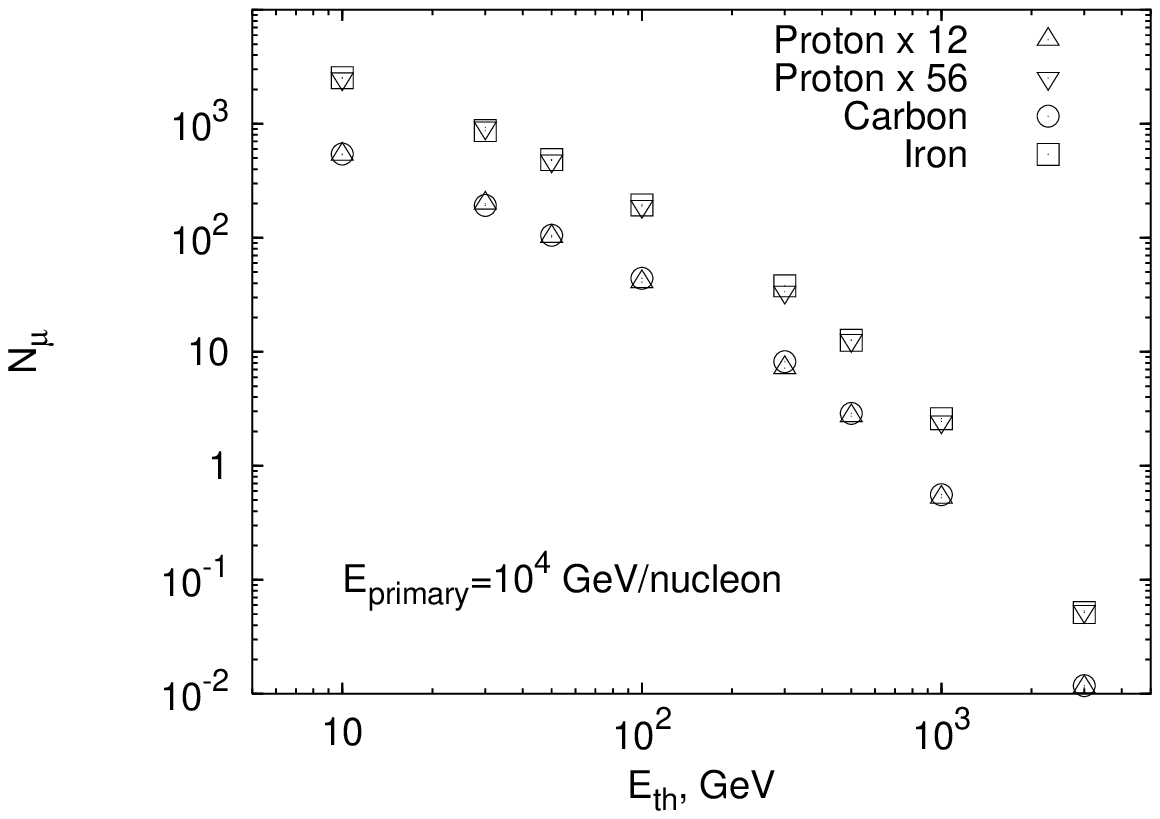}\hfill
\centering\includegraphics[width=.49\textwidth]{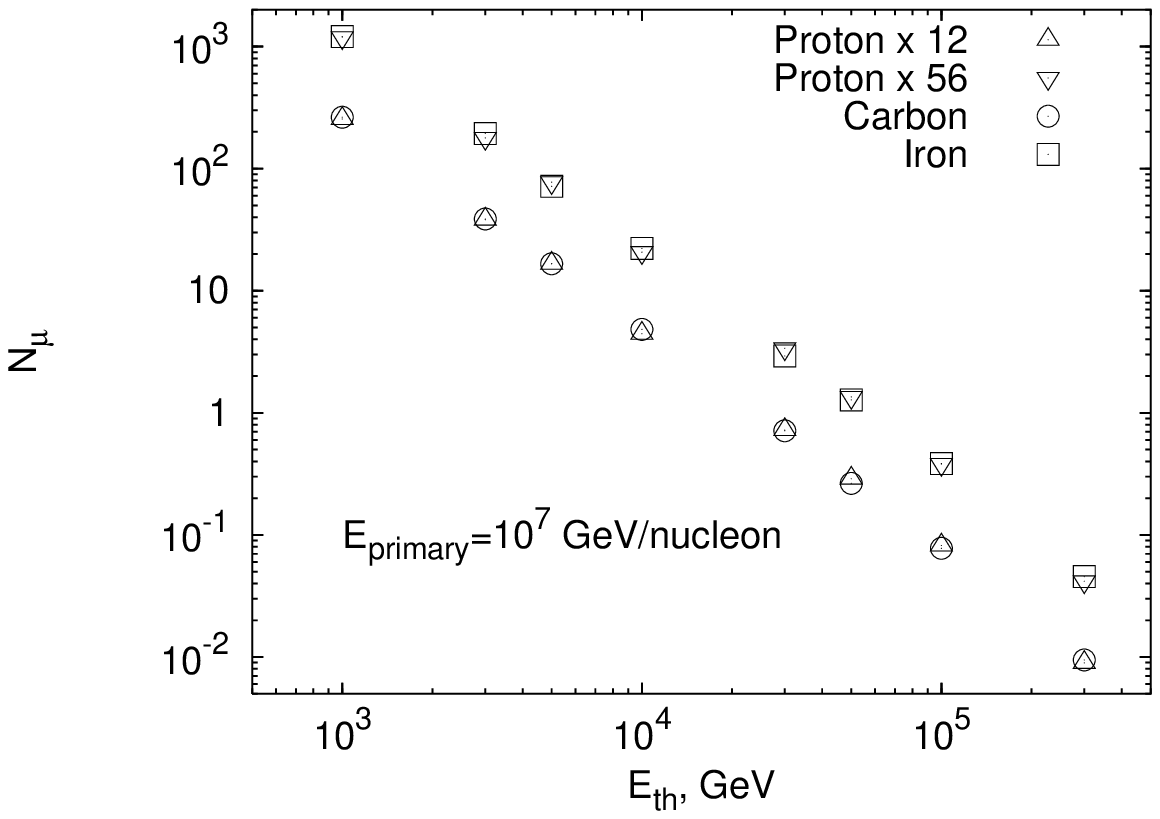}
\caption{Numbers of muons at sea level in cascades, initiated by
primary carbon and iron nuclei. Proton$\times$12 and Proton$\times$56
denote the calculation for carbon and iron nuclei in superposition
model, i.e. the average number of muons in proton initiated shower,
multiplied by 12 and 56 correspondingly. Statistical errors are not
shown, since they are less, than the symbol size.}\label{superp}
\end{figure}

\begin{figure}[h]
\centering\includegraphics[height=.3\textheight]{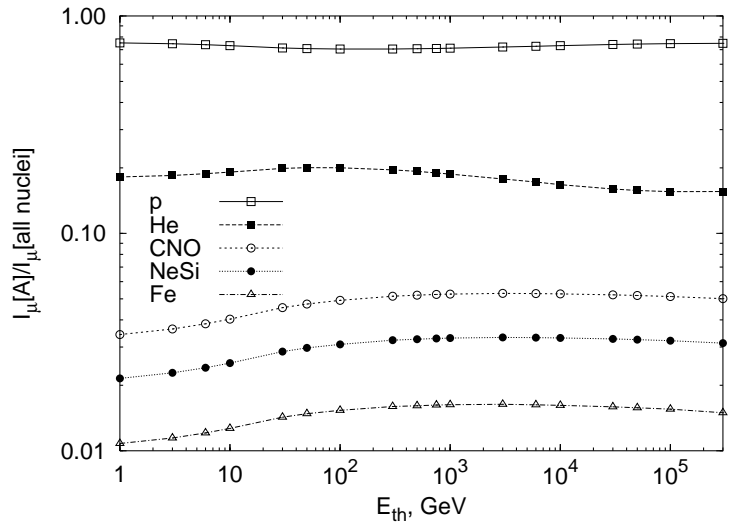}
\caption{Relative contribution of different groups of nuclei
to the integral flux of muons at sea level.}
\label{depos}
\end{figure}

In order to save machine time and accounting, that contribution
of all nuclei, heavier than He to muon spectrum at sea level does
not exceed 7--10\%, we have used a superposition model,
considering nucleus with energy $E_A$ as $A$ nucleons with
energies $E_A/A$. Validity of such approach for calculation of
number of secondaries in EAS at sea level is widely known (see,
{\em e.g.}, \cite{superp}), however, for verification, we have
made our own computations of muon numbers in cascades, initiated
by primary protons, carbon and iron nuclei, for energies
$10^{13}$ and $10^{16}$~eV/n with the QGSJET model. Results are
presented in figure~\ref{superp}. As it is seen, the agreement
between numbers of muons in cascades, simulated with realistic
nuclei fragmentation and those, obtained in the frameworks of
superposition models is excellent. This, as well, allows not to
make a distinction between showers, initiated by protons and
neutrons, and further by nucleons we mean protons. In
figure~\ref{depos} the relative contribution of all
nuclei groups, according to our PCR mass composition, to the
integral muon flux at sea level is shown. Since the contribution of He nuclei
amounts to $\sim20\%$ throughout the entire energy range, thus
mentioned in the previous section $\pm50\%$ spread of
experimental results on helium flux around values, taken in our
model, brings approximately $\pm10\%$ uncertainty to muon flux at
sea level. From this point of view, the uncertainty due to
ambiguity of the experimental situation for heavier nuclei is
insignificant.

To obtain easily the muon spectrum for any model of the primary spectrum,
it is appropriate to perform calculations of integral muon flux in
the following way:
\begin{equation}
I_\mu(>E_\mathrm{th})=\int\limits_{E_\mathrm{th}}^{E_\mathrm{max}}
N_\mu(E_N,>E_\mathrm{th})J_N(E_N)dE_N.
\label{eq_imu}
\end{equation}
Here $N_\mu(E_N,>E_\mathrm{th})$ is an average number of muons
with energy $>E_\mathrm{th}$ in shower from primary nucleon with
energy $E_N$, $J_N(E_N)$~---~differential primary spectrum
converted to spectrum in nucleons, ${E_\mathrm{max}}$~---~energy,
which provides the calculation accuracy $I_\mu(>E_\mathrm{th})\sim
0,1\%$. Most of results, presented below, obtained at
${E_\mathrm{max}}/{E_\mathrm{th}}=3\cdot10^4$.

\begin{figure}
\centering\includegraphics[width=.49\textwidth]{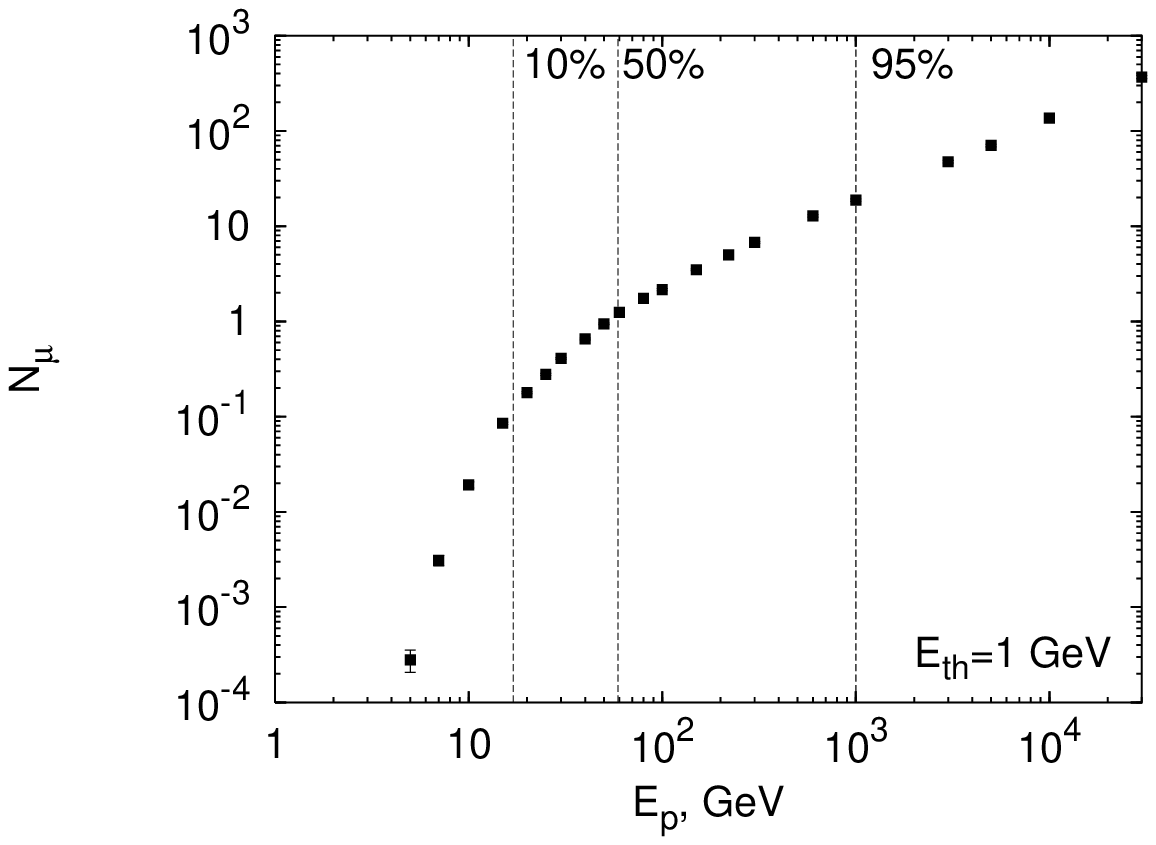} \hfill
\centering\includegraphics[width=.49\textwidth]{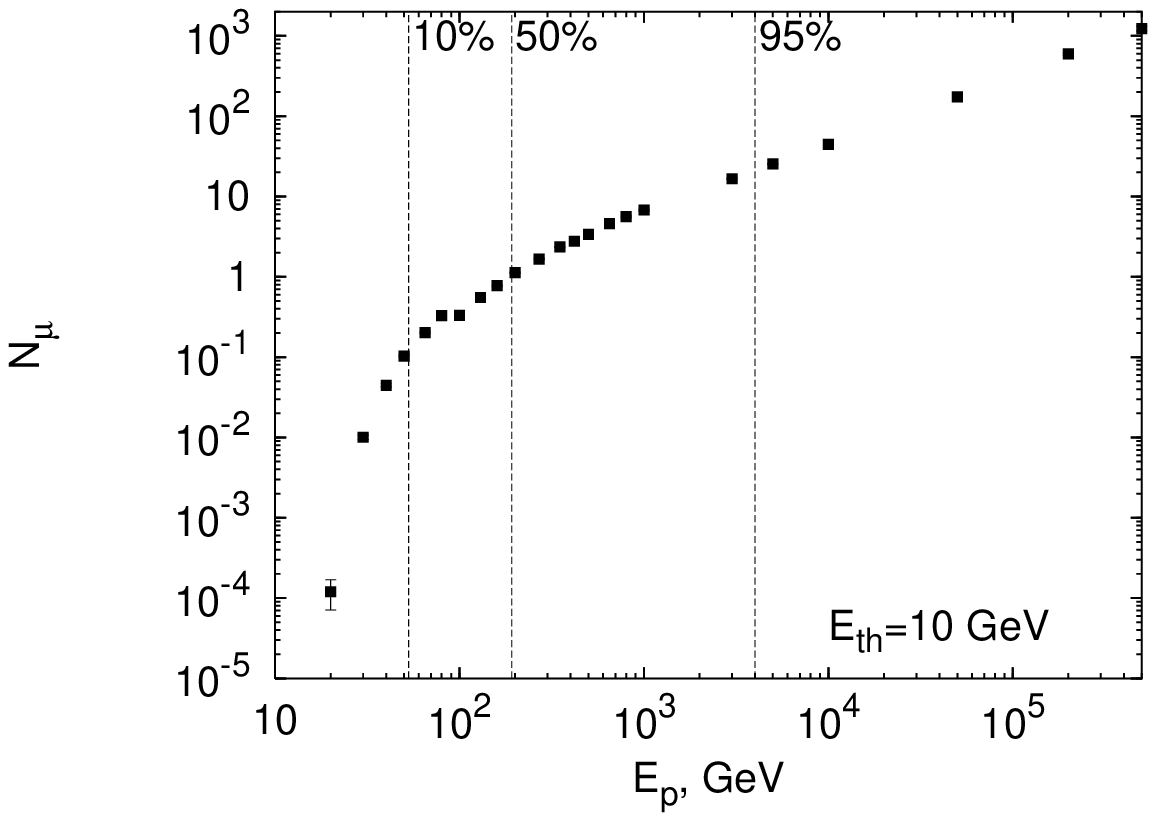}
\centering\includegraphics[width=.49\textwidth]{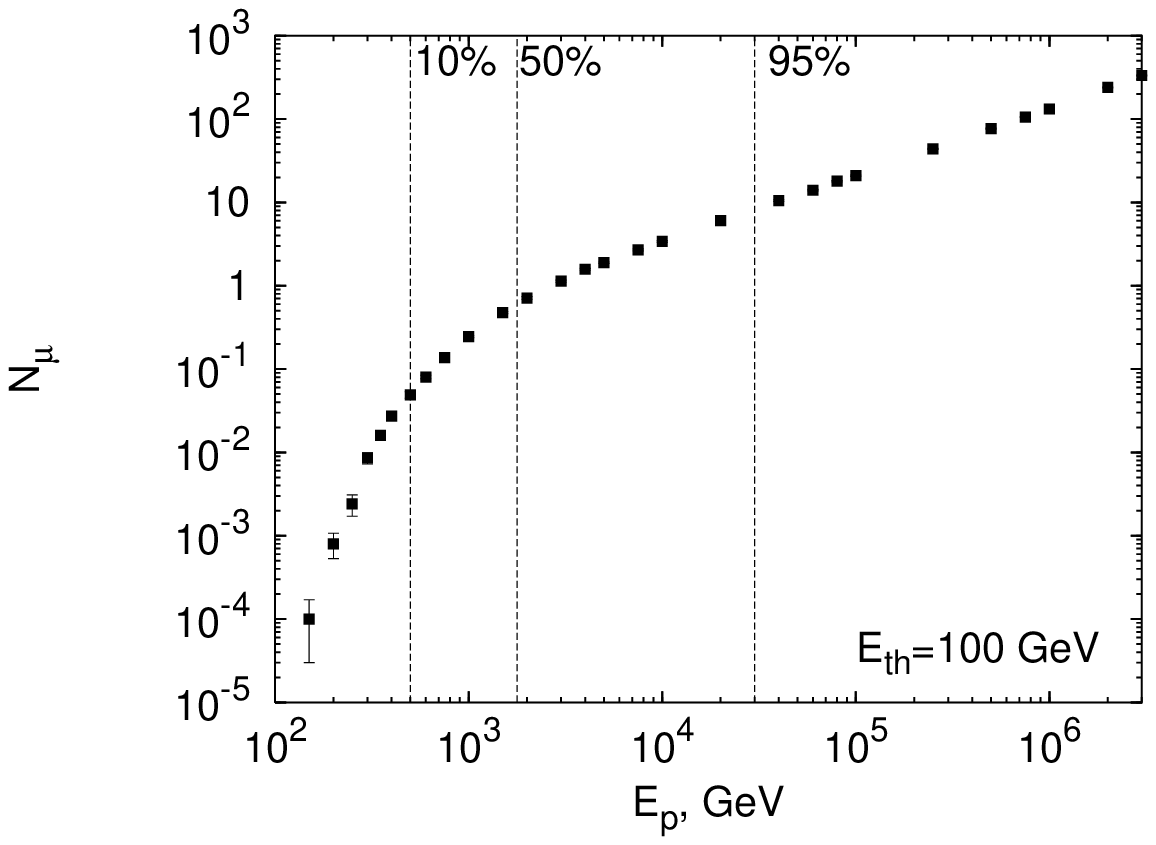}
\hfill \centering\includegraphics[width=.49\textwidth]{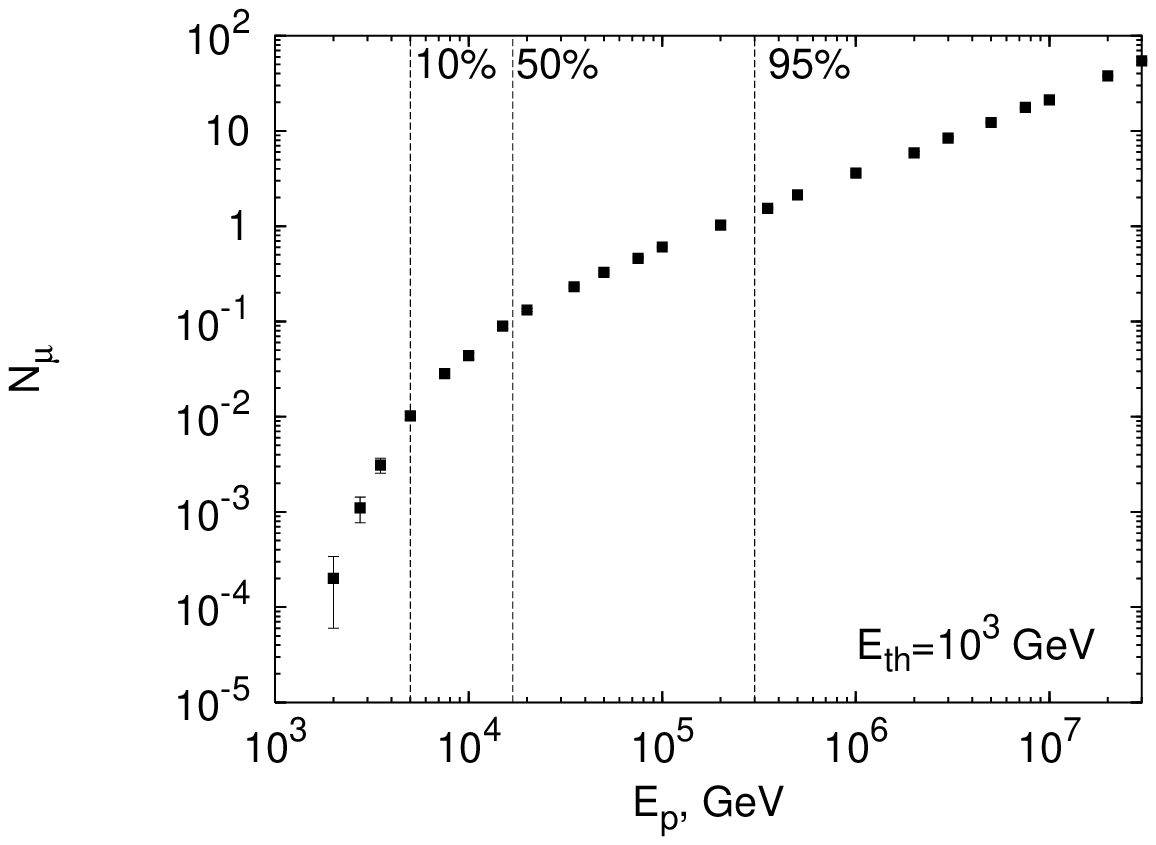}
\centering\includegraphics[width=.49\textwidth]{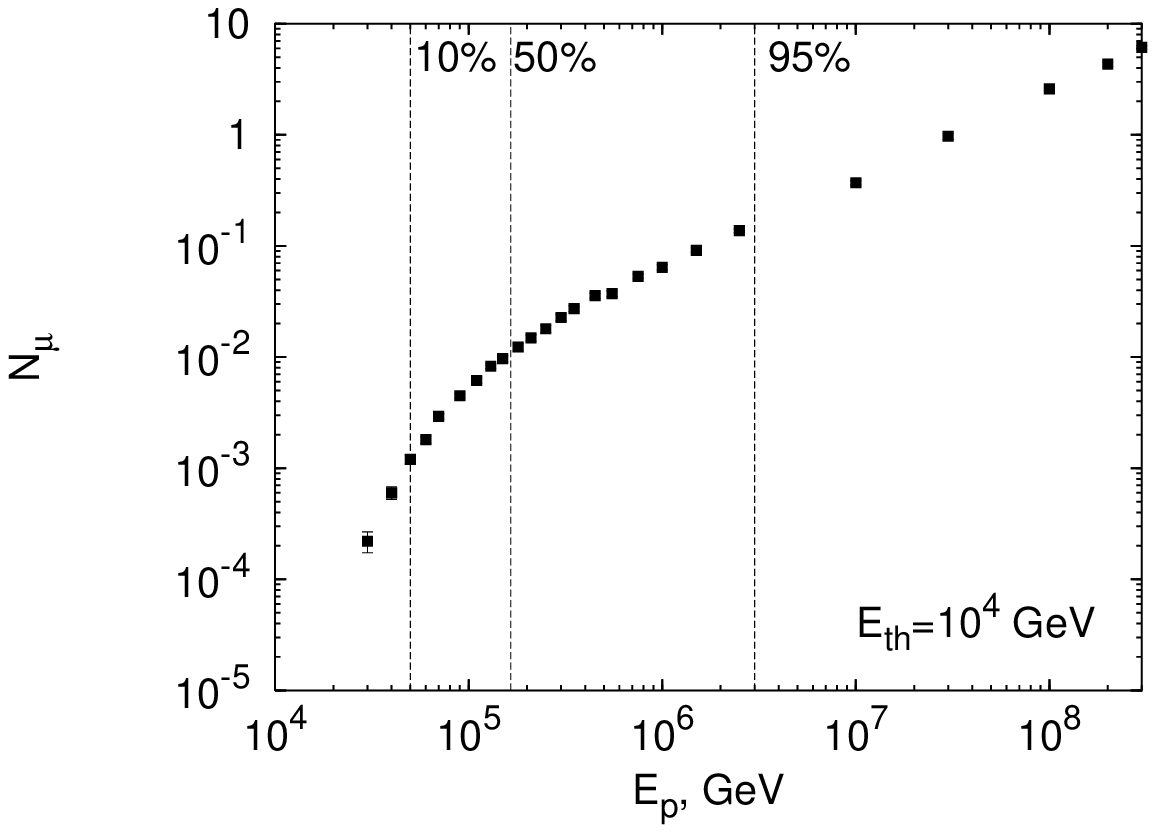} \hfill
\centering\includegraphics[width=.49\textwidth]{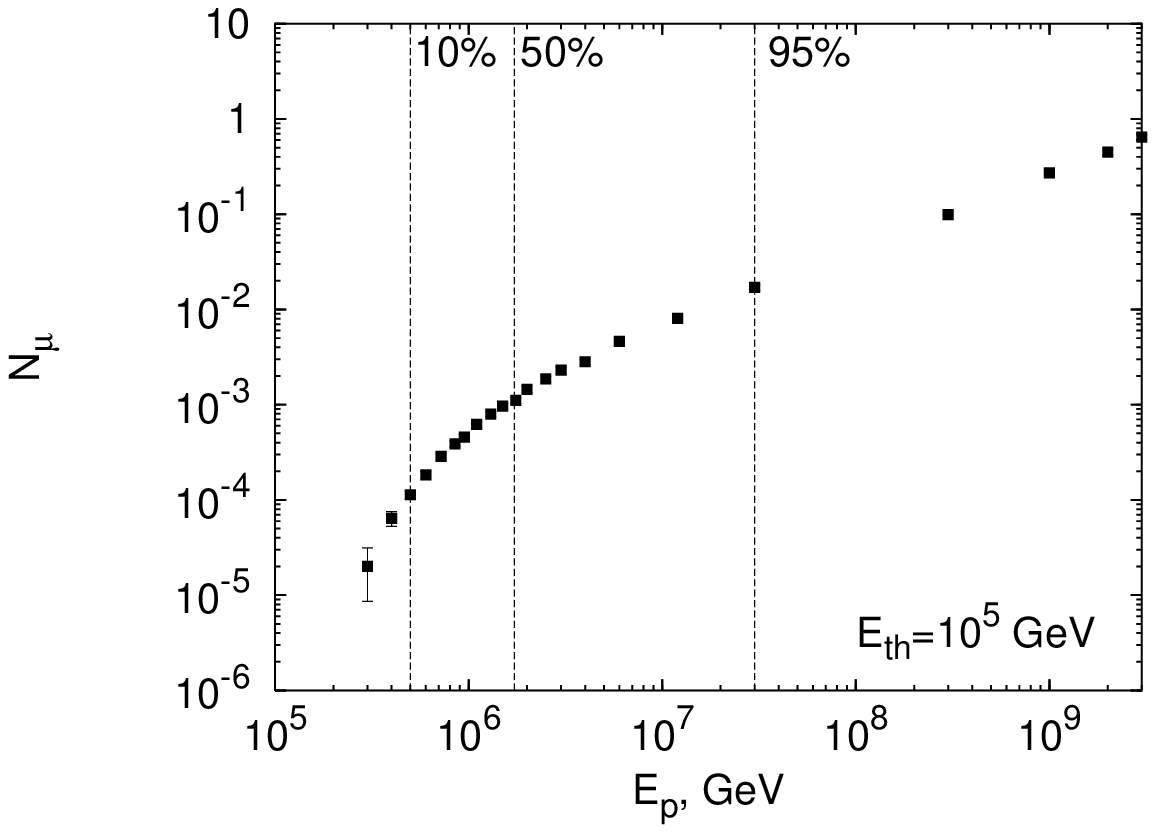}
\caption{Number of muons $N_\mu(E_p,>E_\mathrm{th})$ with energy
above threshold, generated in shower from primary proton with energy
$E_p$. Vertical lines show the areas of primary energies, giving
10\%, 50\% and 95\% contributions to the integral muon flux in case
of power primary spectrum $J_N(E)=1.9\cdot10^4E^{-2.65}$~
$(\mathrm{GeV}\cdot\mathrm{m}^2\cdot\mathrm{sec}\cdot\mathrm{sr})^{-1}$.}
\label{nmu}
\end{figure}

\begin{figure}
\centering\includegraphics[width=.49\textwidth]{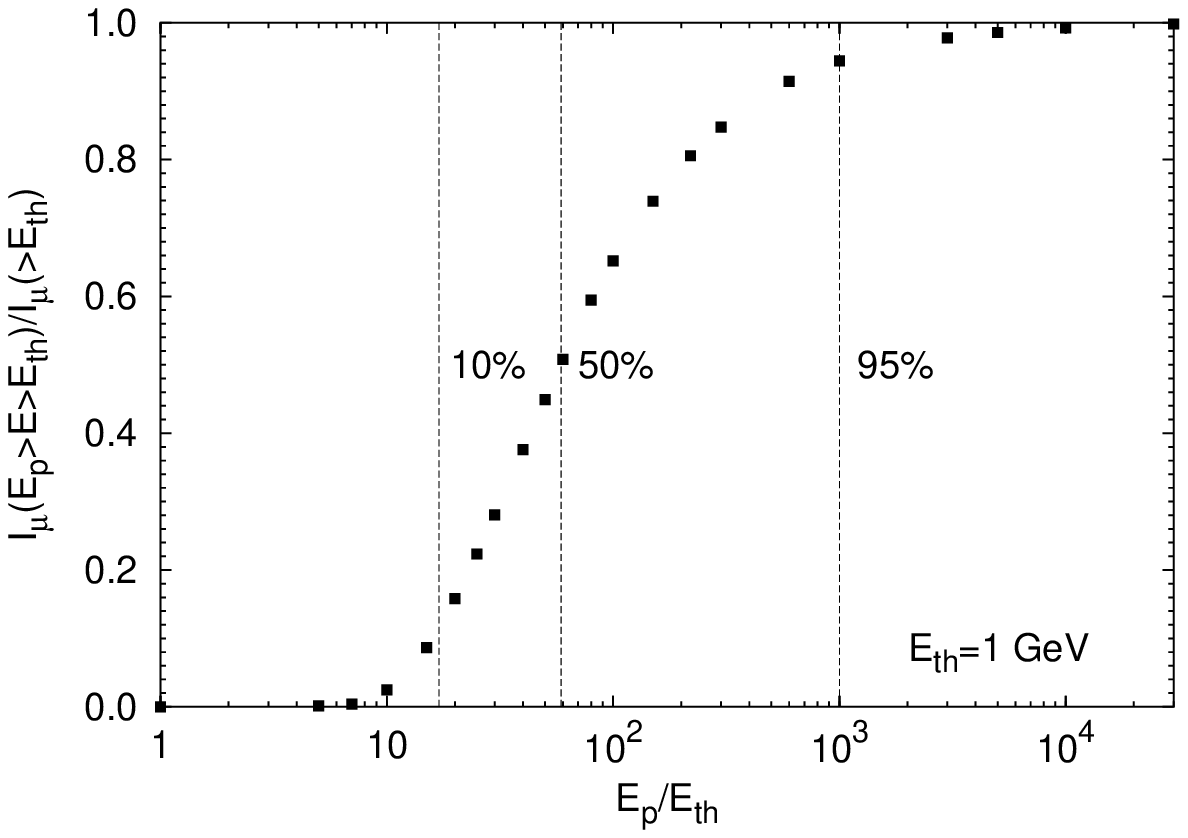}
\hfill
\centering\includegraphics[width=.49\textwidth]{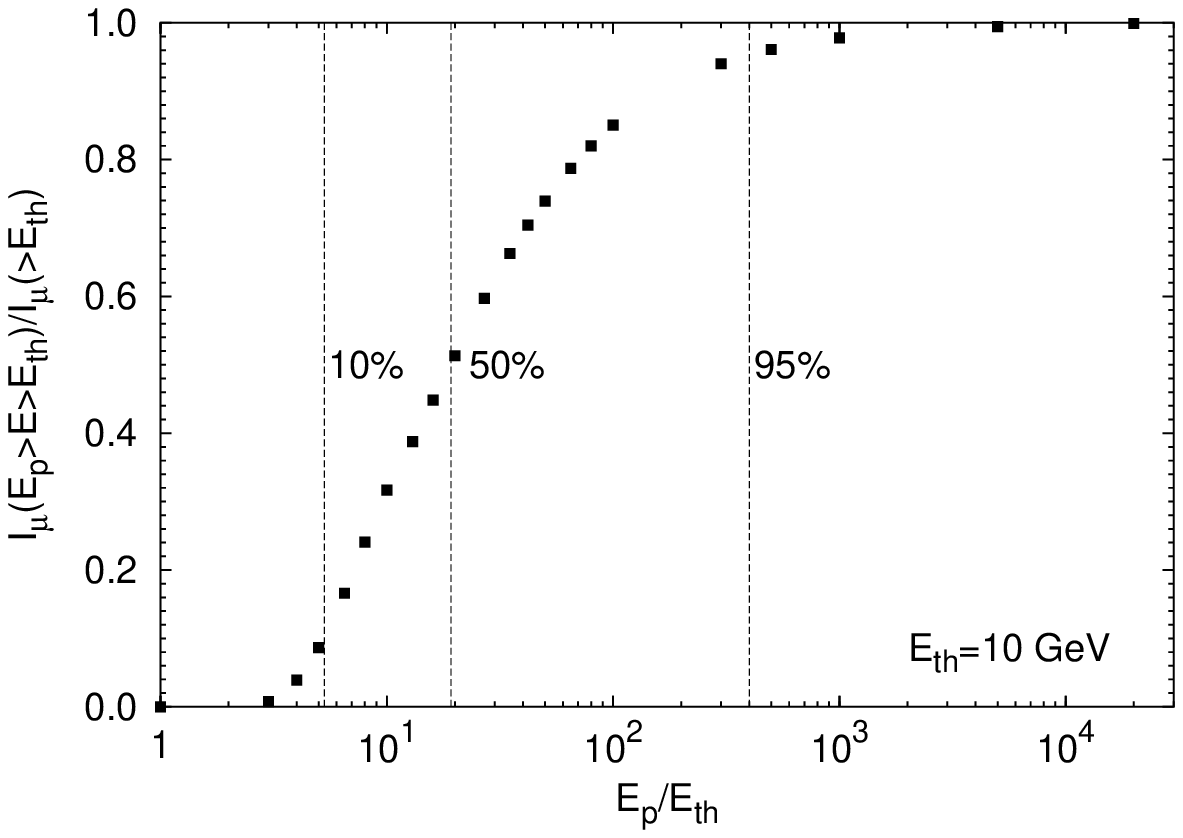}
\centering\includegraphics[width=.49\textwidth]{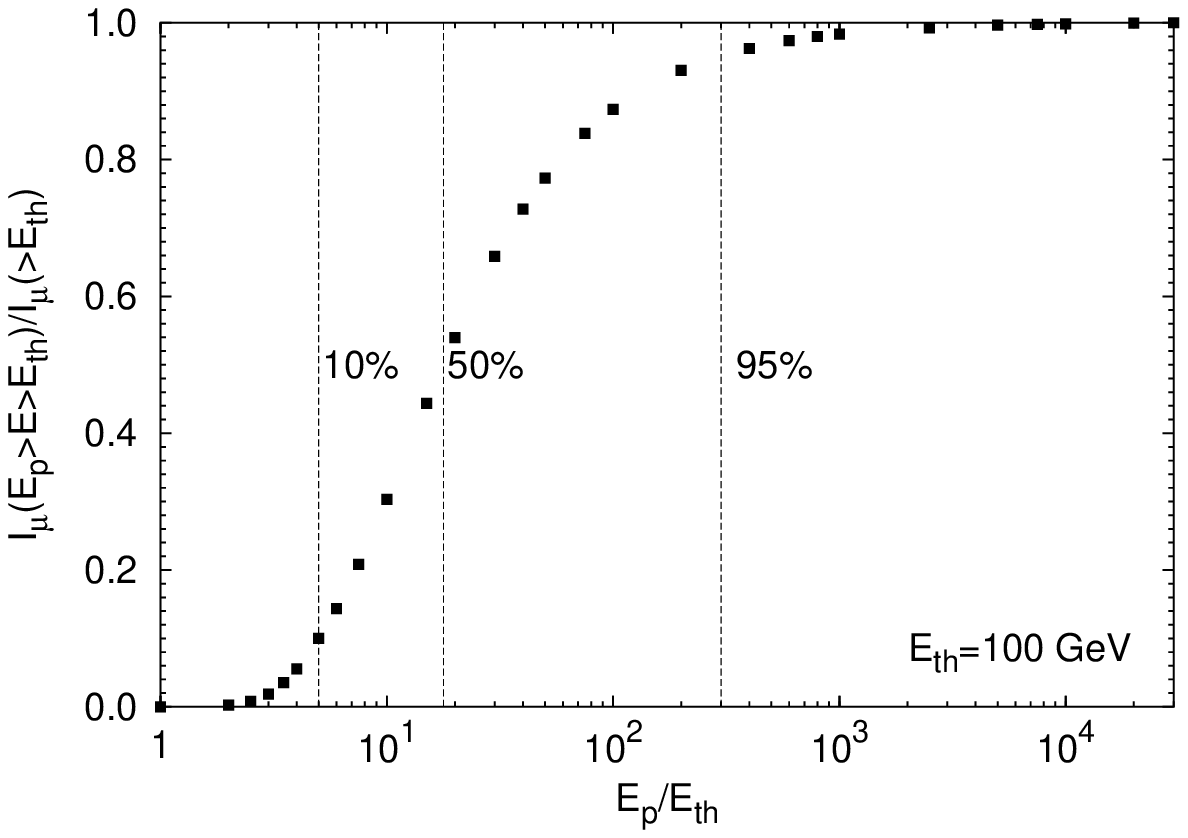}
\hfill
\centering\includegraphics[width=.49\textwidth]{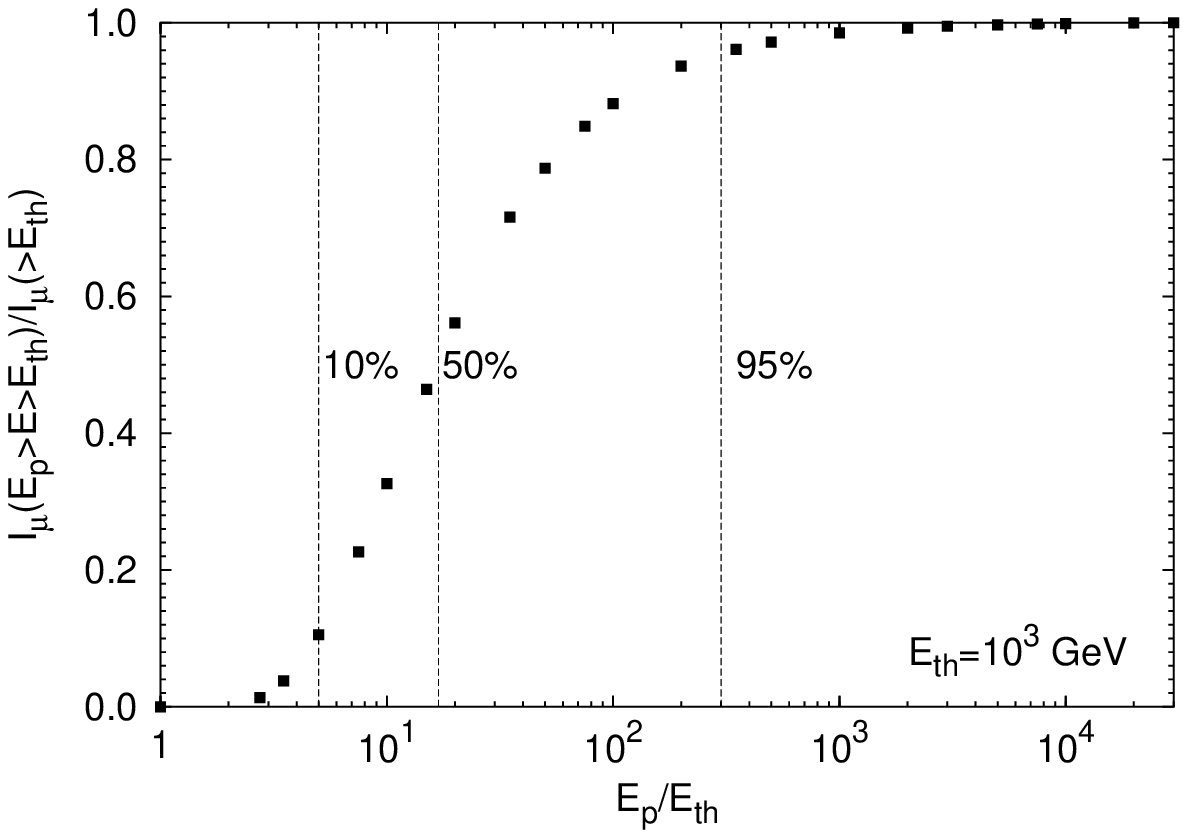}
\centering\includegraphics[width=.49\textwidth]{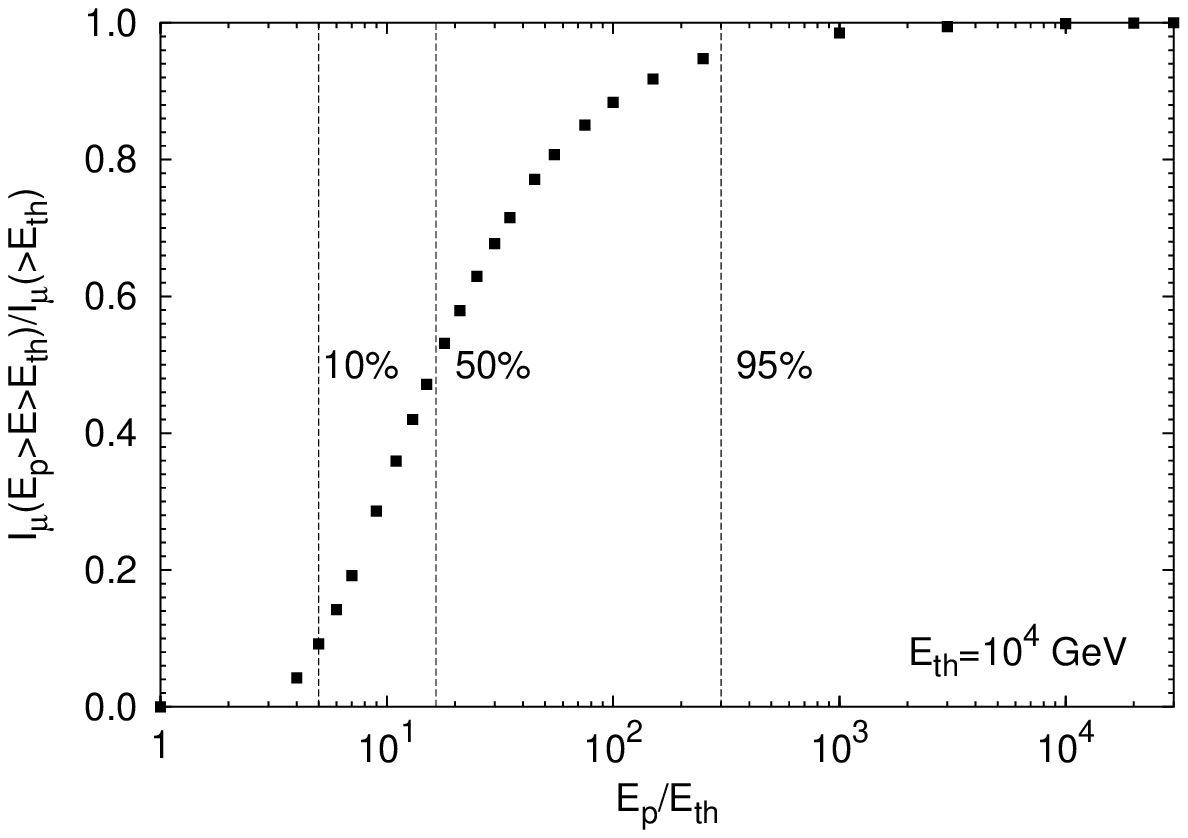}
\hfill
\centering\includegraphics[width=.49\textwidth]{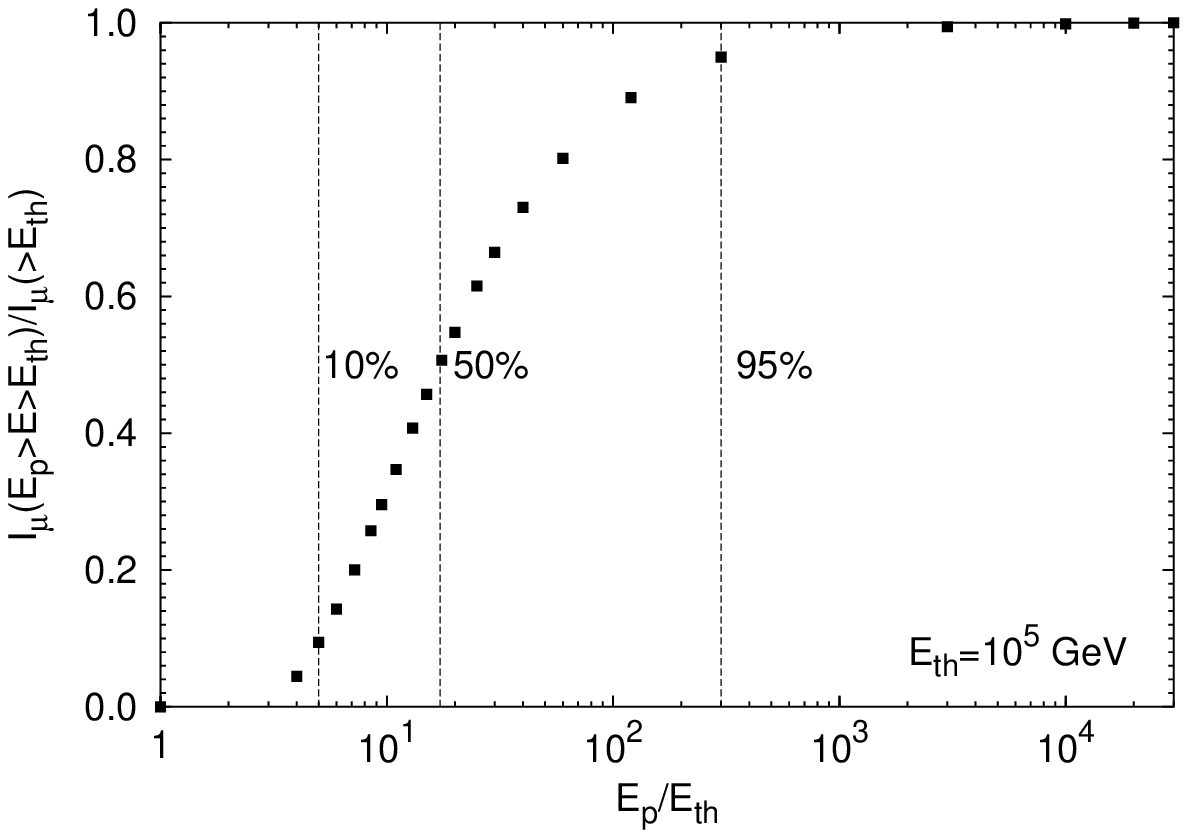}
\caption{Relative contribution of primary energy region from $E_{\th}$ to $E_p$
to the muon integral flux at sea level $I_\mu(>E_\th)$ versus
$E_p/E_{\th}$. Vertical lines have the same meaning as in
figure~\ref{nmu}.}
\label{imu}
\end{figure}

For each threshold energy average muon numbers were computed for
20--25 different primary energies (see figures~\ref{nmu},\,\ref{imu})
with accuracy, generally better than 5\% . Vertical lines in these
figures show the areas of primary energies, giving 10\%, 50\% and
95\% contributions to the integral muon flux in case of power primary
spectrum $J_N(E)=1.9\cdot10^4E^{-2.65}$~
$(\mathrm{GeV}\cdot\mathrm{m}^2\cdot\mathrm{sec}\cdot\mathrm{sr})^{-1}$.
It is seen, that primaries with energies within
$E_\mathrm{th}-300E_\mathrm{th}$ on 95\% determine muon intensity at
sea level. Totally, with the use of the QGSJET it was simulated about
$3\cdot10^7$ showers. Besides, for examination of sensitivity of muon
spectrum to hadron-nucleus interaction model, series of calculations
with the VENUS model have been carried out.

\begin{table}
\label{models}
\caption{Average number of muons with energy
above threshold in shower from primary proton.}
\begin{indented}\item[]
\begin{tabular}{@{}*6{l}}
\multicolumn{6}{c}{Primary energy $E_p=\unit[10^{5}$]{GeV}}\\
\br
\multicolumn{2}{l}{Threshold energy $E_{\th}$}&$1$~GeV&$10$~GeV&$10^2$~GeV&$10^3$~GeV\\
\mr
&\multicolumn{1}{c}{This paper} &1011 &318  &21.0 &0.605\\
\multicolumn{1}{l}{QGSJET}&\multicolumn{1}{c}{\cite{fzka5828}}&1085&316&20.5&0.588\\
&\multicolumn{1}{c}{\cite{knapp97}}&1008&310&20.9&0.696\\
\mr
\multicolumn{1}{l}{VENUS}&\multicolumn{1}{c}{This paper}&1079&347&23.5&0.679\\
&\multicolumn{1}{c}{\cite{fzka5828}}&1150&349&24.0&0.604\\
\br
\end{tabular}
\end{indented}

\vspace*{2mm}
\begin{indented}\item[]
\begin{tabular}{@{}*6{l}}
\multicolumn{6}{c}{Primary energy $E_p=\unit[10^{6}$]{GeV}}\\
\br
\multicolumn{2}{l}{Threshold energy $E_{\th}$}&$1$~GeV&$10$~GeV&$10^2$~GeV&$10^3$~GeV\\
\midrule
&\multicolumn{1}{c}{This paper}&8307&2292&132.3&3.612\\
\multicolumn{1}{l}{QGSJET}&\multicolumn{1}{c}{\cite{fzka5828}}&8298&2207&124.9&3.240\\
&\multicolumn{1}{c}{\cite{knapp97}}&8059&2257&129.8&3.332\\
\mr
\multicolumn{1}{l}{VENUS}&\multicolumn{1}{c}{This paper}&9629&2590&153.5&3.932\\
&\multicolumn{1}{c}{\cite{fzka5828}}&9706&2653&155.0&4.156\\
\br
\end{tabular}
\end{indented}
\end{table}


For the verification of our results, we have performed a comparison
of numbers of muons in showers from primary protons, obtained by
us, with those from papers of CORSIKA
authors~\cite{fzka5828,knapp97} for two primary energies
$E_p=10^5,\,10^6$~GeV (table~1). On the whole, we see good
agreement between the data, presented in this table. Some
difference in muon numbers for $E_\th=10^3$~GeV may be explained
by a statistical error, since simulation of 500 showers, made in the
above mentioned papers for each primary energy, provides
approximately 8\% and 4\% mean square deviation for energies
$E_p=10^5$ and $10^6$~GeV correspondingly.

\section{Results and discussion}

\begin{figure}
\centering\includegraphics[height=0.4\textheight]{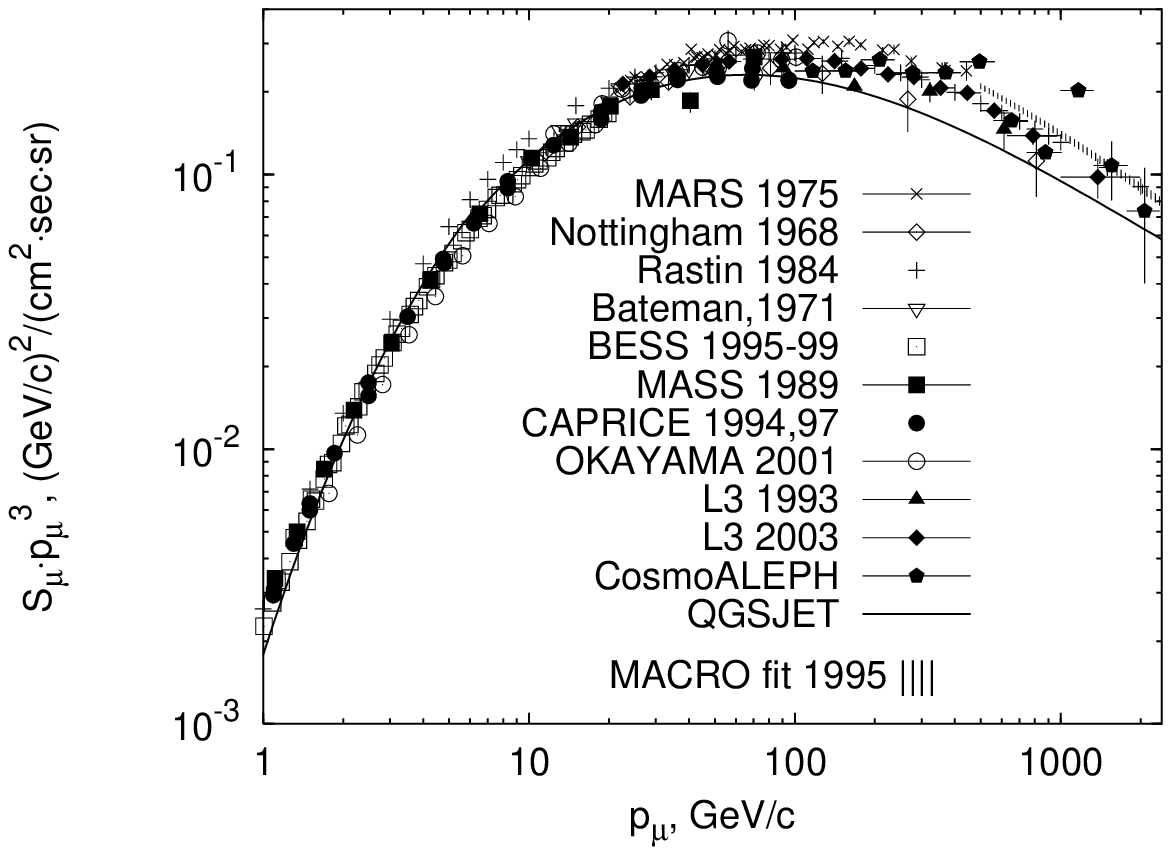}\hfill
\centering\includegraphics[height=0.4\textheight]{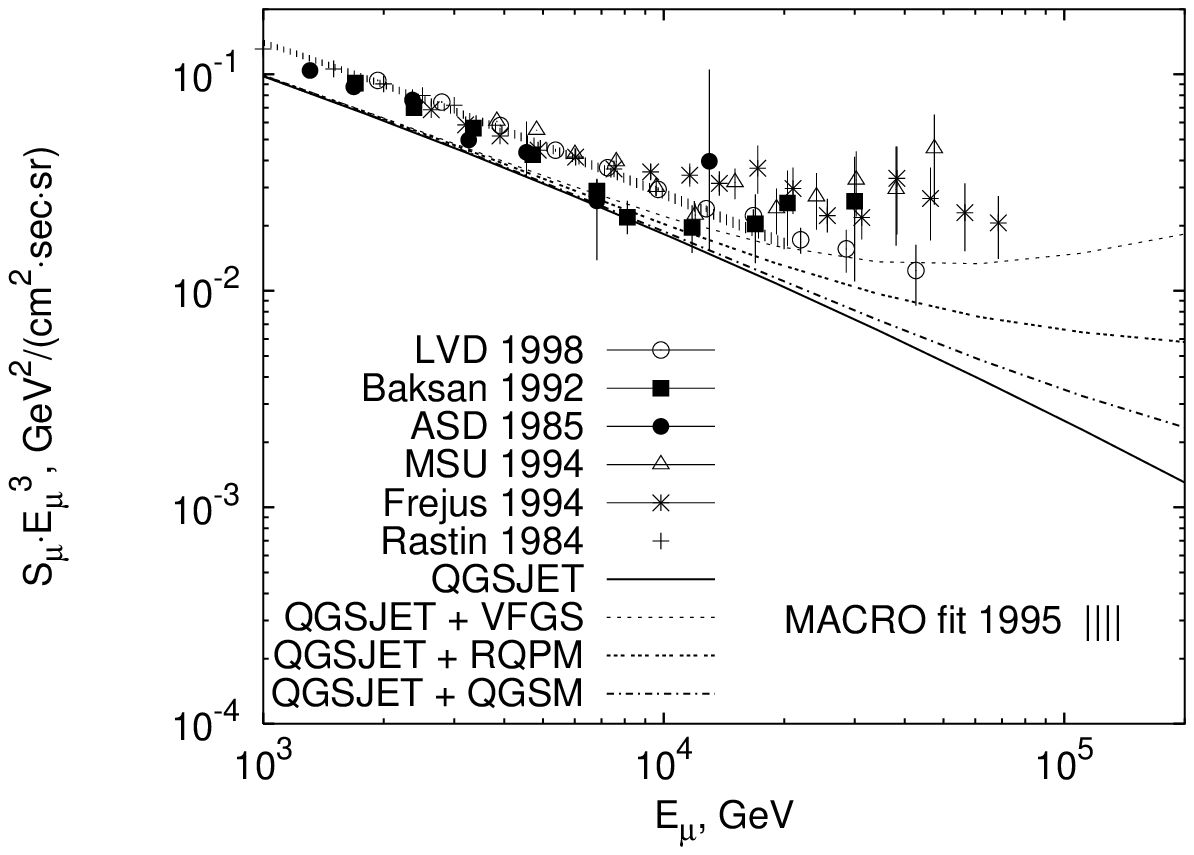}
\caption{Differential muon spectrum at sea level. Experimental data:
\cite{mars75}~MARS~1975, \cite{notting68}~Nottingham~1968,
\cite{rastin}~Rastin~1984, \cite{mass93}~MASS~1989,
\cite{caprice_mu}~CAPRICE~1994,97, \cite{tsuji}~OKAYAMA 2001,
\cite{bateman}~Bateman~1971, \cite{bess_mu}~BESS~1995,~97,~99,
\cite{l3} L3~1993, \cite{l3_03}~L3~2003, \cite{aleph}~CosmoALEPH,
\cite{lvd}~LVD~1998, \cite{baksan92}~Baksan~1992,
\cite{asd}~ASD~1985, \cite{msu_mu}~MSU~1994,
\cite{frejus}~Frejus~1994, \cite{macro}~MACRO fit 1995. QGSJET and
VENUS are the present work calculations with the corresponding
interaction models. Prompt muons spectra: \cite{qgs}~---~RQPM
(recombination quark-parton model), QGSM (quark-gluon string model);
\cite{vfgs}~---~VFGS (Volkova, Fulgione, Galeotti and Saavedra
model).} \label{diff}
\end{figure}

\begin{figure}
\centering\includegraphics[height=0.4\textheight]{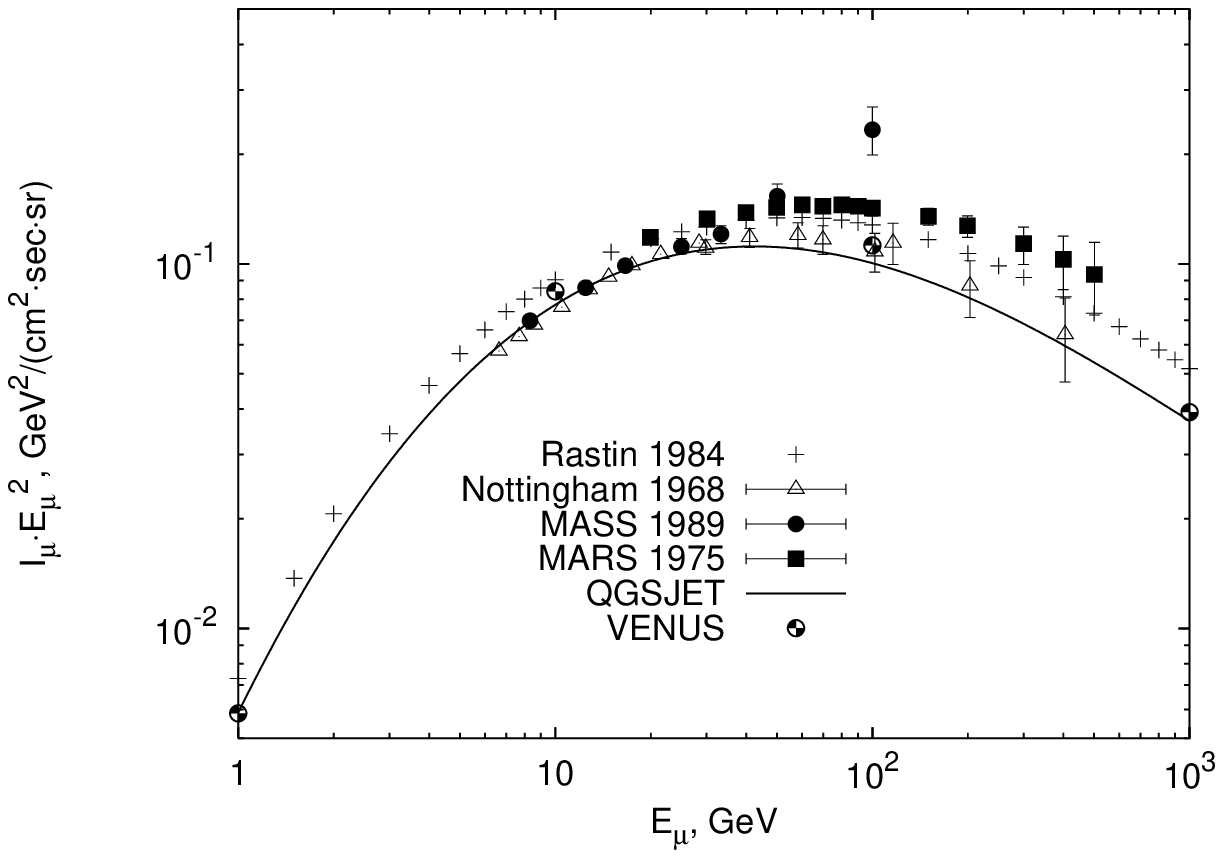}\hfill
\centering\includegraphics[height=0.4\textheight]{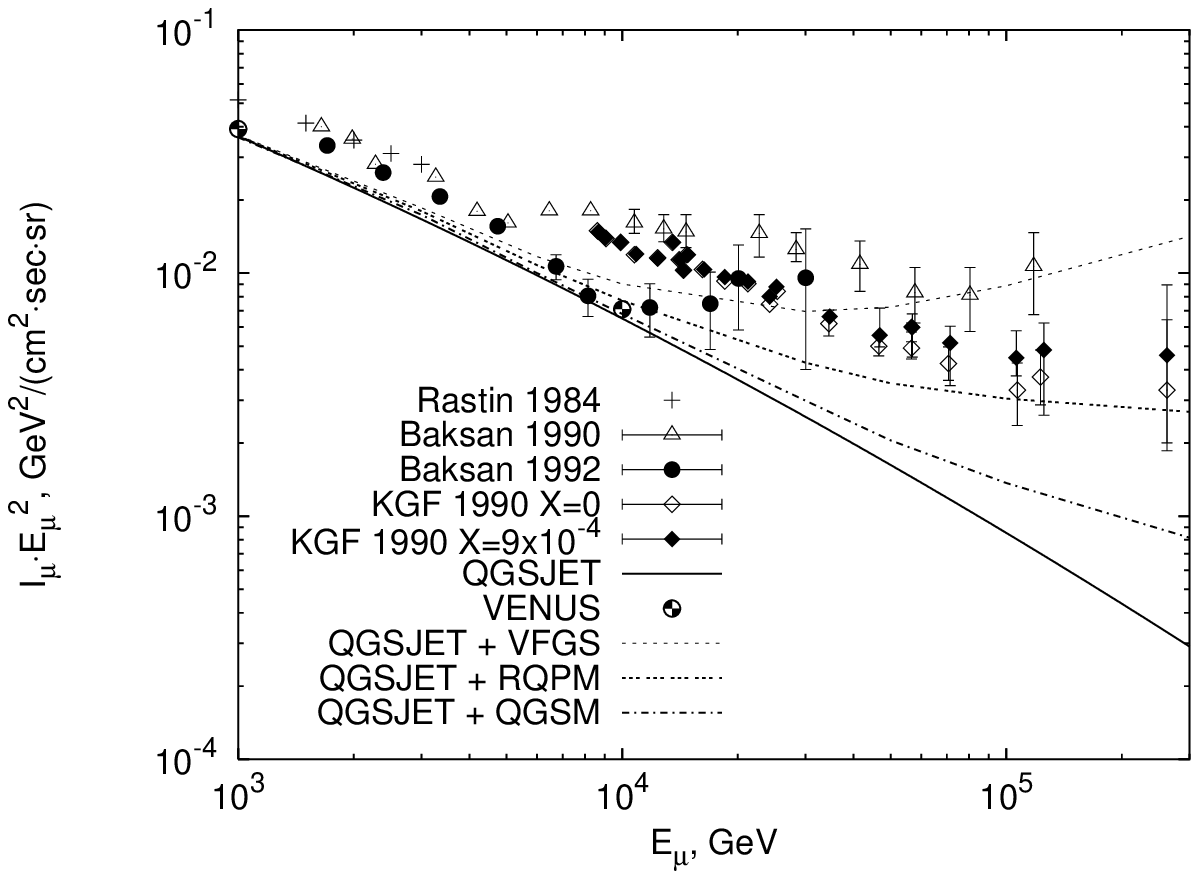}
\caption{Integral muon spectrum at sea level. Experimental data:
\cite{rastin}~Rastin~1984, \cite{notting68}~Nottingham~1968,
\cite{mass93}~MASS~1989, \cite{mars75}~MARS~1975,
\cite{kgf90_1,kgf90_2}~KGF~1990,
\cite{baksan90,baksan92}~Baksan~1990,~1992. QGSJET and VENUS are the
present work calculations with the corresponding interaction models.
Prompt muon spectra: \cite{qgs}~---~RQPM, QGSM;
\cite{vfgs}~---~VFGS.} \label{int}
\end{figure}

Obtained integral and differential muon spectra for our model of
primary spectrum are presented in figures~\ref{diff},~\ref{int}.
It is appropriate to split the consideration of the situation
into three energy intervals.

\textbf{\boldmath{$E_\mu\in[1-10^2$]}~GeV.} In this region our
calculation of differential muon spectrum is in a good agreement with
the most recent and accurate data, obtained by
BESS~1995,97-99~\cite{bess_mu} and CAPRICE~1994,
1997~\cite{caprice_mu}. This, along with the availability of
unambiguous data on primary spectrum and on behaviour of hadronic
cross sections, once again evidences in favour of correctness of the
applied calculation procedure. Selection of BESS and CAPRICE
experiments is motivated by the fact, that measured there muon
spectra have small statistic and systematic errors and excellently
correlate with each other. In these experiments measurements were
made with superconducting magnet spectrometers. Earlier data,
obtained with iron magnet spectrometers differ from each other not
only in values, but also in shapes of measured spectra, that may be
due to the influence of some improperly accounted systematic errors.
More detailed discussion of this question may be found, for example,
in~\cite{summary,bess_mu,tsuji,rastin}.

Approaching to 100 GeV, the calculated muon flux becomes deficient,
conforming only to the CAPRICE data. Two circumstances in this
connection are worth noting. The first is that already from these
energies information on muon intensity is not definite enough. The
second is that muons with $E_\mu\gtrsim100$~GeV are most effectively
produced by the interactions of primaries with
$E_\mathrm{PCR}>1$~TeV/n, measured in space and balloon emulsion
chamber experiments. To this moment we shall pay more attention in
section~5.

\textbf{\boldmath{$E_\mu\in[10^2-10^4$]}~GeV.} Experimental data on
differential proton spectrum for corresponding primary energies
$10^3-10^6$~GeV of SOKOL, MUBEE, JACEE and RUNJOB show good
interconsistency, but from $10^5$~GeV they become less definite for
technical and natural (low flux) reasons. Experimental data on muon
component have smaller errors and agree within $\sim20\%$. In this
region along with direct measurements on the surface
(MARS~\cite{mars75}, Nottingham~\cite{notting68,rastin},
L3~\cite{l3,l3_03}) there are results obtained at underground
installations, in which the sea level muon spectrum is reconstructed
from the `depth-intensity' curve and in other ways. The latter are
the data of the Baksan neutrino observatory
(BNO)~\cite{baksan90,baksan92}, Artyomovsk scintillation
detector~\cite{asd}, MSU~\cite{msu_mu}, Frejus~\cite{frejus},
LVD~\cite{lvd}, MACRO~\cite{macro}, KGF~\cite{kgf90_1,kgf90_2} and
CosmoALEPH~\cite{aleph}. Most of them are grouped within fit limits,
given by MACRO, and comprise a well correlated data set. Comparison
of the obtained differential and integral muon spectra with these
data does not support reasonable expectation `PCR data
fit$\to$nuclear cascades$\to$muon data fit'. The deviation of our
results from the data of `MACRO zone' is about 30--40\%. This fact
requires revising all steps of our calculation in order to realize
how real the problem is and what may be done to overcome it. First of
all, one should analyze an adequacy of nucleon cascades description.
As it is shown in the previous section, our computations do not
contain methodical errors, and total statistical + interpolation +
integration error is less, than 5\%. From figure~\ref{int} it is also
seen, that the maximal difference $<10\%$ in predictions of the
QGSJET and VENUS hadronic interaction models is also practically out
of significance in this situation. Since the VENUS model was
initially chosen by us as the model, providing one of the largest
muon numbers in shower (see \cite{fzka5828}), hence 10\% should be
regarded as the utmost possible increase of muon intensity for any
interaction model in comparison with the QGSJET.

\begin{figure}
\centering\includegraphics[width=.5\textwidth]{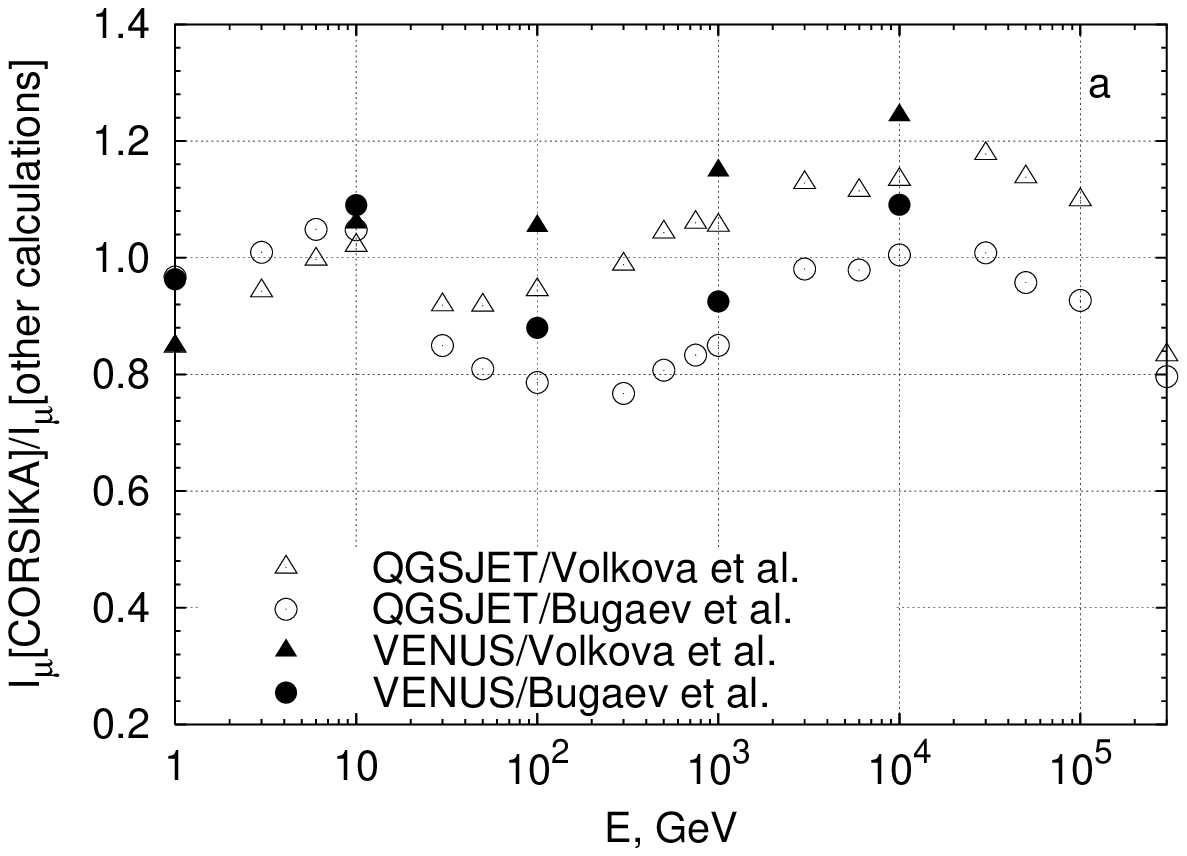}\hfill
\centering\includegraphics[width=.5\textwidth]{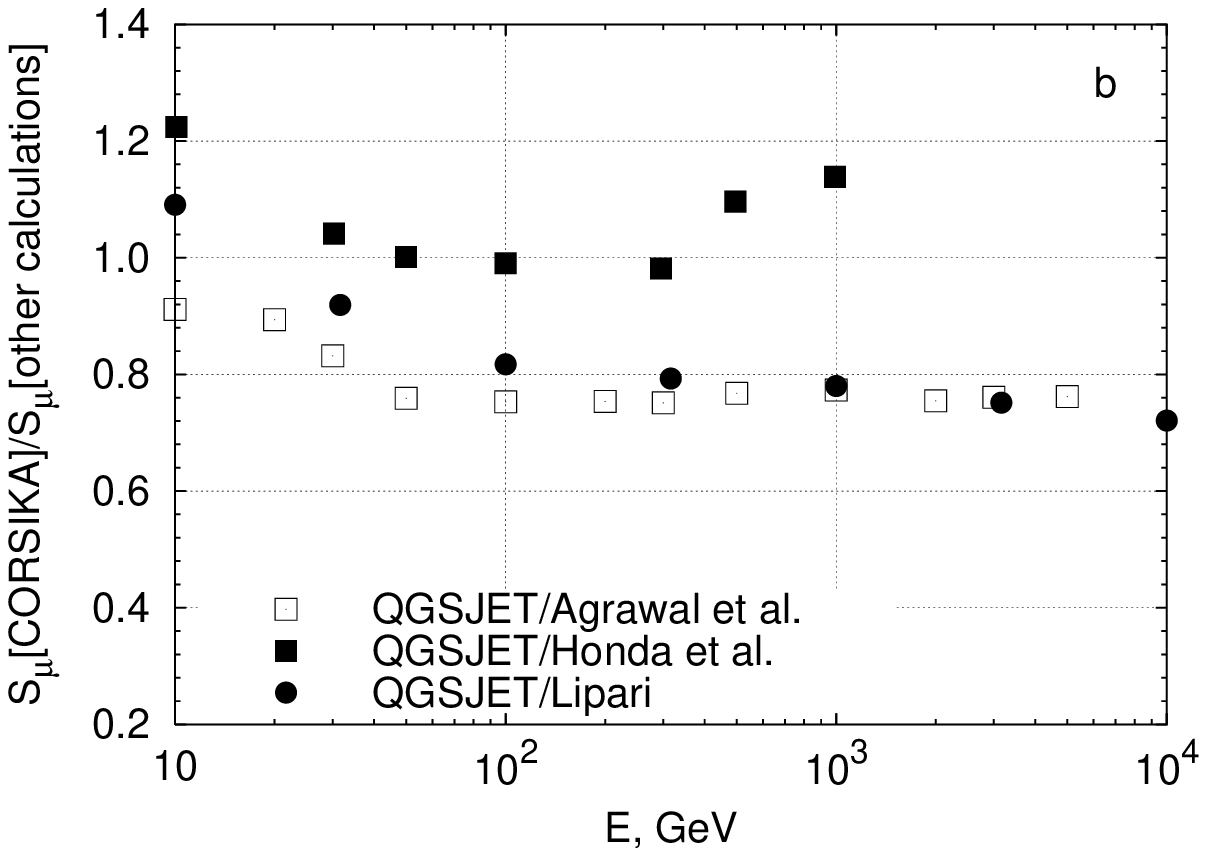}
\caption{Ratio of integral (a) and differential (b) muon fluxes,
obtained with CORSIKA, to the muon fluxes calculated in
papers~\cite{volk}\,(Volkova et al.), \cite{bugaev}\,(Bugaev et al.),
\cite{agrawal}~Agrawal et al., \cite{honda}~Honda et al.,
\cite{lipari}~Lipari. The CORSIKA results are for the same PCR
spectra as used in these papers.} \label{comp}
\end{figure}

In this connection the results of the earlier works, not indicating
any problems with the description of the muon data, deserve some
closer consideration. We analyze here five
works~\cite{volk,lipari,honda,agrawal,bugaev}. The collation of the
PCR spectra with the experimental data, performed in section~2,
allows to split these papers into 2 groups: papers with
excessive~\cite{volk,honda,bugaev} and
appropriate~\cite{lipari,agrawal} fluxes. In order to reveal
differences in hadronic interactions and calculation technique, we
have made computations of muon fluxes for the same PCR spectra, as
used in the discussed works. Results are presented in
figure~\ref{comp}. From this comparison it becomes evident, why no
muon deficit was discovered earlier. In addition to overvalued
nucleon fluxes, in the papers from the first group the muon yield is
larger or, at least, comparable with that of QGSJET and VENUS. Muon
production in the papers with appropriate primary fluxes is
$20\%-30\%$ higher, than obtained by us, and this difference is
sufficient to mask the muon shortage. Besides, PCR flux from the
work~\cite{agrawal} becomes excessive from $\sim10$~TeV, as we have
already pointed in section~2. The data in respect of muon production
of the latter two papers closely agree, apparently because in both of
these works essentially the same authors participated. In
\cite{agrawal} also an explanation of relatively large muon yield may
be found. In this calculations, for simulation of hadronic
interactions TARGET code had been applied, but with the enhancement
of pion and kaon production. This resulted in increase of
contribution of kaon decays to muon flux with $E_\mu>1$~TeV up to a
very large value of 50\%. Of course, uncertainties in description of
hadronic interaction prevail, but, as our calculations show,
discrepancies in muon yield at sea level between the widely
recognized and extensively tested models, included in CORSIKA, are at
the level of $\pm10\%$. These models are evidently free from many
drawbacks, peculiar to the models from the discussed works, and more
thoroughly developed. Hence, likely the largest part of the
discrepancy should be attributed to underestimation of primary
nucleon flux, but not to the incorrectness of the simulation of the
nuclear cascades in the atmosphere.

There is no need to repeat arguments from section~2 in favour of
adequacy our PCR model, since it may be regarded just as a valid
representation of the current experimental situation. We should only
repeat, that our proton spectrum does not present a fit of the
experimental data for $E_p=10^4-10^5$~GeV, but even on $\sim$20\%
overestimates them. This was done to smooth the discussed problem. If
directly input in calculations fits of primary H and He spectra,
obtained by JACEE, with addition of 10\% contribution of heavier
nuclei, or `standard' spectrum, proposed by Gaisser and Honda, then
resulting integral muon flux will be even lower, than ours (see
table~2).

There is a possibility to compensate some of the muon deficit by
increase of intensity of heavier nuclei in our model. But to get 10\%
addition to muon flux it is necessary to rise helium flux 50\% or to
double aggregate flux of CNO, Ne-Si and Fe groups. This mechanism is
rather ineffective, because limitations, set by the data on
all-particle spectrum thus would require to lower fluxes of other,
lighter species. Summarizing, we see, that today for this energy
region the situation is determined rather rigidly and the stated
problem of muon deficit can not be resolved without some
extraordinary assumptions. Necessity in rise of all PCR spectra as
high, as possible to diminish the muon shortage, indicates, that
measurements with emulsion chambers systematically understate fluxes
of the primaries. A large spread in the data on PCR nuclei with
$A\geq4$ provides some more ground for such speculations and
demonstrates necessity in further improvement of experimental
technique. This question is closely related to the appropriate
description of the characteristics of nuclear interactions and in
more details is considered in section~5.

\renewcommand{\tabcolsep}{5pt}
\begin{table}
\label{other_spectra}
\caption{Ratio of the integral muon
intensity, calculated with the use of the present paper PCR model, to
the intensities, obtained with the JACEE PCR fit~\cite{jacee} and the
model, proposed by Gaisser and Honda~\cite{gaisser2002}.}
\begin{indented}\item[]
\begin{tabular}{cccccccccc}
\br &\centre{9}{Muon threshold energies, GeV}\\ \ns &\crule{9}\\
\multicolumn{1}{c}{Ratio}&\multicolumn{1}{c}{$10$}&
\multicolumn{1}{c}{$30$}&\multicolumn{1}{c}{$10^2$}&
\multicolumn{1}{c}{$3\cdot10^2$}&\multicolumn{1}{c}{$10^3$}&
\multicolumn{1}{c}{$3\cdot10^3$}&\multicolumn{1}{c}{$10^4$}&
\multicolumn{1}{c}{$3\cdot10^4$}&\multicolumn{1}{c}{$10^5$}\\ \mr
$I_\mu[\mathrm{present}]/I_\mu$\cite{jacee}&$-$&$-$&0.98&1.07&1.12&1.11&1.04&0.94&0.79\\
$I_\mu[\mathrm{present}]/I_\mu$\cite{gaisser2002}&1.02&1.08&1.12&1.13&1.10&1.03&0.89&0.75&0.58\\
\br
\end{tabular}
\end{indented}
\end{table}

\textbf{\boldmath{$E_\mu\in[10^4-3\cdot10^5$]}~GeV.} Analysis of
this interval in the present situation looks vain for the
following reasons:

i. Obviously, muon deficit takes place for these energies too, and
its nature and quantity are unclear;

ii. There is no experimental data on chemical composition of PCR
spectrum for energies higher than $10^6$~GeV, that is why, as in our
case, one has to use only theoretical guidelines, describing the
origin and transport of cosmic rays in interstellar medium.

iii. Behavior of hadron-nucleus cross-sections for
$E_\mathrm{lab}>10^5$~GeV, despite the decrease of divergence between
predictions of different models, remains also not enough studied,
that results in additional uncertainty in calculations of the muon
spectrum.

iv. All muon component measurements for the given energies are
obtained from results of underground experiments and contain quite
large systematic errors, that causes their poor mutual agreement.
Dominant sources of the mentioned errors are such factors as
incomplete information on chemical composition of overburden rock and
necessity to input in calculation a ratio $X$ of prompt muons to
$\pi,K$-muons as a function of energy, which is not only unknown, but
should be determined from the experimental data.

Taking aforesaid into account, it seems impossible to make any
definite conclusions on behaviour of PCR spectrum and on prompt muon
contribution on the basis of sea level data in this energy region.
Though from the comparison with the experimental data the VFGS
(Volkova, Fulgione, Galeotti and Saavedra) model~\cite{vfgs} of charm
generation looks as the most preferable, however, here it is needless
to stress all unreliability of this deduction.

\section{Possible reasons of systematic PCR flux underestimation}

In this section we intend to discuss some problems, that are inherent
to the study of high-energy cosmic rays spectra with the application
of the emulsion chambers (EC). In addition to all factors, listed
below, one can easily add some more, that experimenters encounter
during processing of the EC data, including, for example, not
negligible experimental bias. We are mostly concentrated at the
factors, depending on the description of nuclear interactions, and in
this aspect the situation may eventually turn out similar to that
with the calculations of sea level muons. The necessity of further
improvements in the experimental technique is clearly recognized in
connection with the serious disagreements between the current data on
different groups of nuclei, and short history of space and balloon EC
experiments gives many examples of how the reported data changed with
the gain of experience and statistics. Further investigations will
inevitably provide information, that improves our knowledge, and,
possibly, will differ from it.

The most detailed description of the experimental technique is
presented by the RUNJOB collaboration~\cite{runjob} and we shall
mostly rely on their data, bearing in mind that the procedure is
generally the same in all EC experiments (MUBEE, JACEE and RUNJOB).
The key moment lies in the determination of the energy, transferred
to the electromagnetic component $\sum E_\gamma$ in the cascades,
initiated by the interaction of a primary particle within EC. This
energy is related to the initial energy $E_0$ via partial
inelasticity coefficient $k$ (we shall denote so $k_\gamma$ to avoid
further confusion with the spectral index $\gamma$). The spectrum of
primaries is obtained from the spectrum of electromagnetic cascades
(EMC) with a simple shift in the energy scale by the value
\[
C^{-1}(k,\gamma)=\left[\int\limits_0^1k^\gamma
f(k)dk\right]^{-1/\gamma},
\]
here $f(k)$ is a distribution function of $k$ and $\gamma$ is the
spectral index of primary spectrum $J(E)\sim E^{-(\gamma+1)}$. The
given statement, at the least dating back to as early as
1962~\cite{babayan}, holds true, provided $f(k)$ does not depend on
$E_0$ in a wide energy range. Usually independence of $C(k,\gamma)$
on energy is argued from the evidence, that total and partial $<k>$
inelasticities are constant. But this may prove to be incorrect (see,
e.g~\cite{grigorov1973}), since $<k^\gamma>$ essentially depends on
the contribution of the large values of $k$ to the distribution
$f(k)$, and even though $<k>$ is constant, `tail' of $f(k)$ may change with
$E_0$. Besides, large body of information on behaviour of the total
inelasticity, presented in the literature, points at its slow growth
with energy. Another important moment, is the difference in $<k>$
between predictions of various interaction models. In comparison with
FRITIOF, applied by RUNJOB, VENUS gives $\sim10\%$ lower value of
partial inelasticitiy $<k>$ in a single {\em p-A}, He{\em-A}
collisions~\cite{runjob} (other experiments do not provide
information neither on the code used, nor on the single interaction
characteristics). Note, that 10\% correction  to the energy
conversion from $\sum E_\gamma$ to $E_0$ for power PCR flux $J(E)\sim
E^{-2.8}$ is equivalent to $\sim30\%$ correction to the intensity.
Thus, at first sight, the use of the VENUS provides not only the
largest number of muons at sea level, but, as well may lead to an
increase of the measured proton and helium fluxes. In fact, the
estimation of the influence of the discussed effect on the results of
EC measurements is much more complicated. Firstly, one needs to
evaluate total energy, transferred to EMC in successive interactions
of secondary particles, i.e. from the whole shower, not from single
interaction, and to obtain distribution function $f(k)$ for this
fraction $k$ of the initial energy. Secondly, this distribution for
$k$ is rather broad, so the conversion to $E_0$ for individual shower
is impossible and it is necessary to derive effective value of
$C(k,\gamma)$, allowing to get PCR spectrum from the EMC one. So,
simple knowledge of differences in $<k>$ in the first interaction
does not allow to make straightforward qualitative estimate of
changes in the final energy shift value. The analysis becomes even
more complicated, if to account, that the choice of the interaction
model strongly influences the previous steps of the experimental data
processing. Namely, it affects the determination of $\sum E_\gamma$
and detection efficiency. For evaluation of both of these values a
complete simulation of nuclear cascades in EC is required. For
example, in the RUNJOB experiment the `actual' EMC energy $\sum
E_{\gamma,true}$ is obtained from the estimated with the $\gamma$-ray
core method one $\sum E_{\gamma,esti}$ via direct Monte-Carlo
simulation of showers with the FRITIOF code. In the JACEE
experiment~\cite{jacee1986,jacee} determination of cascade energies
also involves complete Monte-Carlo calculations of the transition
curves, used for calibration of the $\sum E_\gamma$, derived from the
direct electron counting or from the X-ray film densitometry.
Analogous calculations are required for evaluation of the detection
efficiency (see detailed description in~\cite{runjob}).

Thus, deviations between the interaction models in hadron-nucleus
cross-sections, energy spectra, multiplicities and phase space
distributions (this is important for estimation  of the most
energetic in the lab system and back-scattered in CMS numbers of
gamma-quanta. VENUS predicts the largest quantity of the latter,
compared to the other models~\cite{fzka5828}) of seconday particles,
in fluctuations of the energy, transferred to EMC, etc., may
significantly influence the interpretation of the EC measurements
results. The mentioned arguments give enough ground for performing of
thorough analysis of sensitivity of the EC data to the use of various
interaction models, which is still missing. This as well would allow
to make fully consistent estimations of the cosmic ray fluxes, i.e.
with the application of the single model, both at the top of the
atmosphere and at the sea level.

Besides, it is possible that some data distortion may come from
changes or anomalies in characteristics of hadronic interactions at
very high-energies. At present, these questions are widely discussed
in connection with the data of EAS experiments, also employing EC
technique.

Another hypothetical reason, which may cause systematic errors in
processing of the EC experiments data, could be the presence of
sizable fraction of unusual component in PCR. This may be regarded
equivalent to the `change' in hadronic interactions and all aforesaid
about that effect holds true in the given case. In this connection,
rising interest to the primary antiproton component deserves closer
attention.

\begin{figure}
\centering\includegraphics[height=.3\textheight]{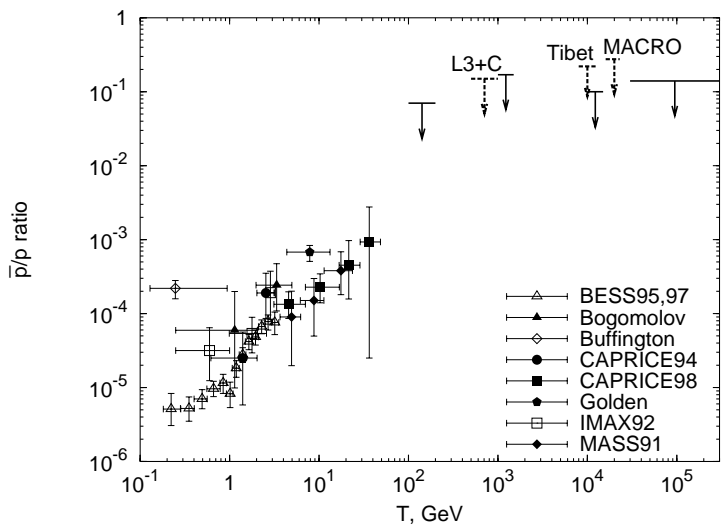}
\caption{The primary antiproton to proton fluxes experimental data.
Direct measurements: \cite{caprice_ap98}~CAPRICE98,
\cite{caprice_ap94}~CAPRICE94, \cite{bess_ap}~BESS95,97,
\cite{imax_ap}~IMAX92, \cite{mass_ap}~MASS91,
\cite{bogomolov}~Bogomolov et al., \cite{buffington}~Buffington et
al., \cite{golden}~Golden et al. Upper estimates of the $\bar{p}/p$
ratio from the ground-based experiments: \cite{stephens}~---~solid
arrows; dashed arrows: \cite{tibet}~Tibet, \cite{l3_antip}~L3+C,
\cite{MACRO_antip}~MACRO.} \label{antip}
\end{figure}

As it is appropriately supposed, antiprotons present natural compound
of PCR, produced mostly in interactions of nuclei (dominantly
protons) with interstellar medium. Other, exotic $\bar{p}$ generation
mechanisms, result in softer spectra and they are out of interest for
the energies, we are dealing with (for more information, see
references in caption to figure~\ref{antip} and {\em
e.g.}~\cite{moskal03}). The experimental study of $\bar{p}$ component
began in 1970's from the first balloon measurements (Golden et
al.~\cite{golden} and Bogomolov et al.~\cite{bogomolov}), and today
the direct data cover region from few MeV to approximately 40~GeV
(figure~\ref{antip}). Higher energy region is scarcely studied and is
marked by few indirect upper estimates of $\bar{p}/p$ ratio. One of
them by Stephens (1985)~\cite{stephens} is derived from the analysis
of sea level muon charge ratio. Soon we shall present elsewhere such
calculations with the employment of improved since then knowledge on
PCR fluxes, hadronic interactions and muon charge ratio. Other three
$\bar{p}/p$ estimates by Tibet~\cite{tibet}, L3+C~\cite{l3_antip} and
MACRO~\cite{MACRO_antip} collaborations are made with the use of the
Moon and the Sun shadow effects. The idea of this approach is the
following: positively and negatively charged particles are deflected
by the Earth's magnetic field to the opposite sides, and when the Sun
or the Moon, during their transit over detector, block particles,
deficit of secondary particles should be observed, if antimatter
presents in PCR along with the matter, at the both opposite sides of
the real Sun/Moon positions. Since these experiments failed to detect
`antimatter shadow', they could report only upper limits of
$\bar{p}/p$ ratio to be equal 22\% at 10~TeV~\cite{tibet}, 27.5\% at
20~TeV~\cite{MACRO_antip} and 15\% for muon sample with
$E_\mu>65$~GeV (primary energy $\sim1$~TeV)~\cite{l3_antip}.
Theoretical predictions are not able to clarify the situation, since
there is no generally recognized propagation theory, reproducing
satisfactory even low energy data on $\bar{p}/p$ ratio together with
all totality of the data on elemental nuclei abundances, secondary to
primary nuclei ratios, and fluxes of positrons and gammas. Standard
diffusion/reacceleration models fail to do this, requiring
introduction of specific assumptions, generally working for
description of some kind of observations, but leading to
contradictions in other cases. Among such assumptions one may readily
call necessity in `artificial breaks in diffusion coefficients and in
the primary injection spectrum'~\cite{moskal03} to match both
$\bar{p}/p$ and secondary to primary nuclei ratios, suggestion about
harder interstellar proton and/or electron spectra to explain well
known `GeV excess' in diffuse gamma ray flux, etc. (for more
information see, {\em e.g.}~\cite{moskal03,moskal00,mori,pohl} and
references therein). The detailed discussion of these problems is
beyond the scope of our paper and we shall return to it elsewhere.

In any case, it is clear, that there is no firm ground to state
neither the presence of significant fraction of $\bar{p}$'s for high
energies, nor the reverse and that actual $\bar{p}/p$ ratio is still
to be determined via direct and various kinds of indirect
observations. Thus, one can not completely exclude probability, that
among events, identified in EC as protons, sizable fraction of
antiprotons presents. As it is already said above, it is difficult to
estimate the influence of the differences between {\em p-A} and {\em
$\bar{p}$-A} interaction characteristics on the results of PCR
measurements with EC. It is known that antiprotons deposit more
energy into $\gamma$'s, than $p$ in a single
interaction~\cite{fzka5828}, but, we should repeat, the complete
analysis of changes, produced in the measured $p+\bar p(?)$ flux
requires to perform evaluation of many parameters, that is possible
to do only via direct Monte-Carlo simulations with respect to the
specific experimental conditions.

We have made similar estimates for the total muon flux at sea level.
As we have elucidated, actually all primary nucleons in our
calculations are protons. Replacing them by antiprotons, we have
computed integral muon flux for
$E_\mathrm{th}=10^2,\,10^3,\,10^4$~GeV and have not found any
noticeable changes, in comparison with the calculations for protons.
In other words, total vertical sea level muon flux turned out to be
insensitive to the fraction of antiprotons in PCR and, possibly, the
same may happen with the EC results. Of course, this is not true for
$\mu^+$ and $\mu^-$ spectra, and $\mu^+/\mu^-$ ratio, their
calculations may help to determine presence of $\bar{p}$'s in PCR.

\begin{figure}
\centering\includegraphics[height=.3\textheight]{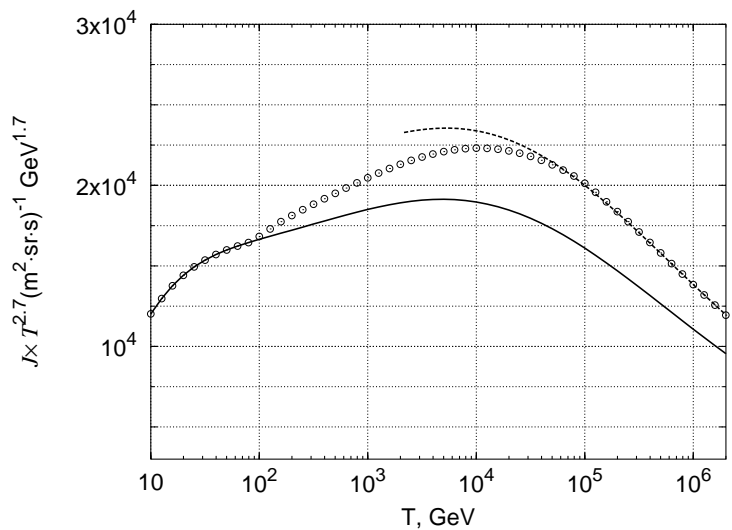}
\caption{Initially applied nucleon spectrum (solid line) and its part
for the region, relevant to the EC measurements, 8\% shifted in the
energy (dashed line). Circles show the spectrum, used to get the
lower estimate of the sea level muon flux, presented in
figures~\ref{adiff},\,\ref{aint}.} \label{nucleon_shift}
\end{figure}

\begin{figure}
\centering\includegraphics[height=0.4\textheight]{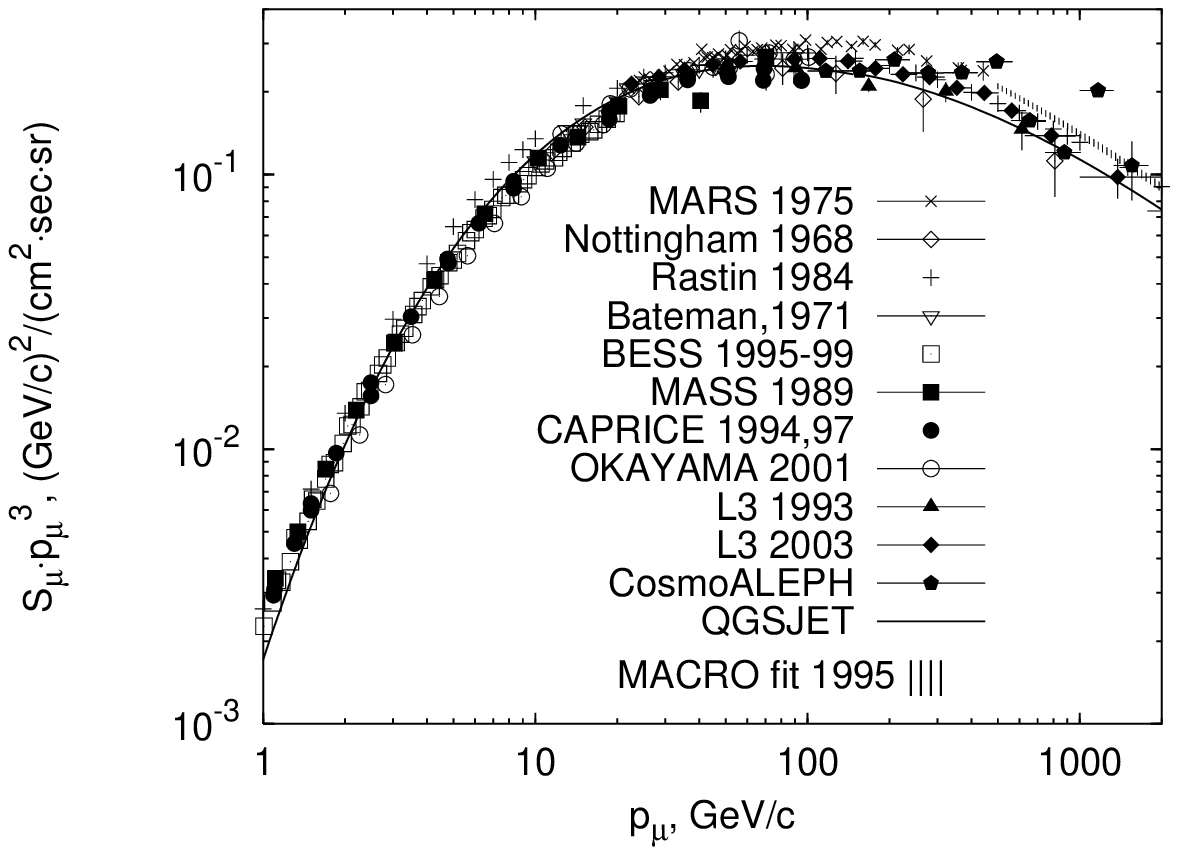}\hfill
\centering\includegraphics[height=0.4\textheight]{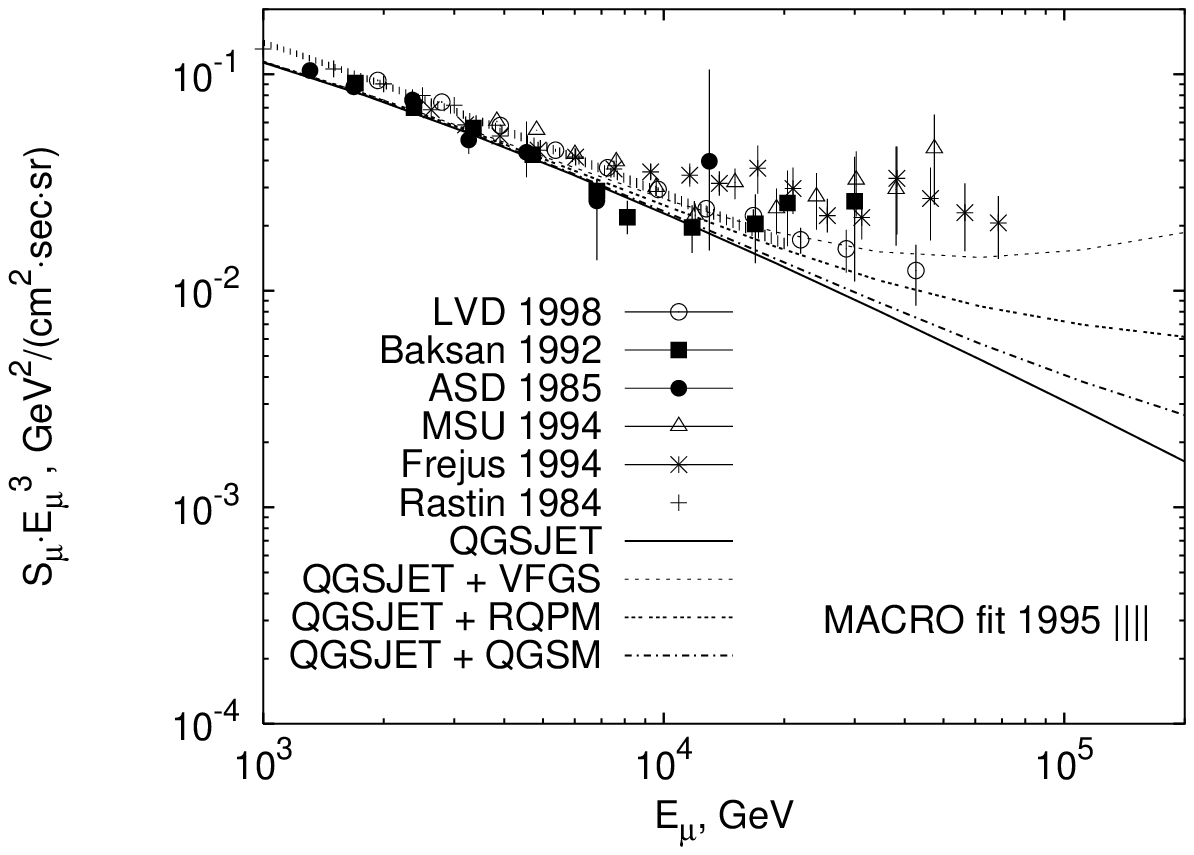}
\caption{Differential muon spectrum at sea level. QGSJET and
VENUS~---~present work calculations with the corresponding
interaction models for the primary nucleon spectrum, shown in
figure~\ref{nucleon_shift} with open circles. Other designations are
the same as in figure~\ref{diff}.} \label{adiff}
\end{figure}

\begin{figure}
\centering\includegraphics[height=0.4\textheight]{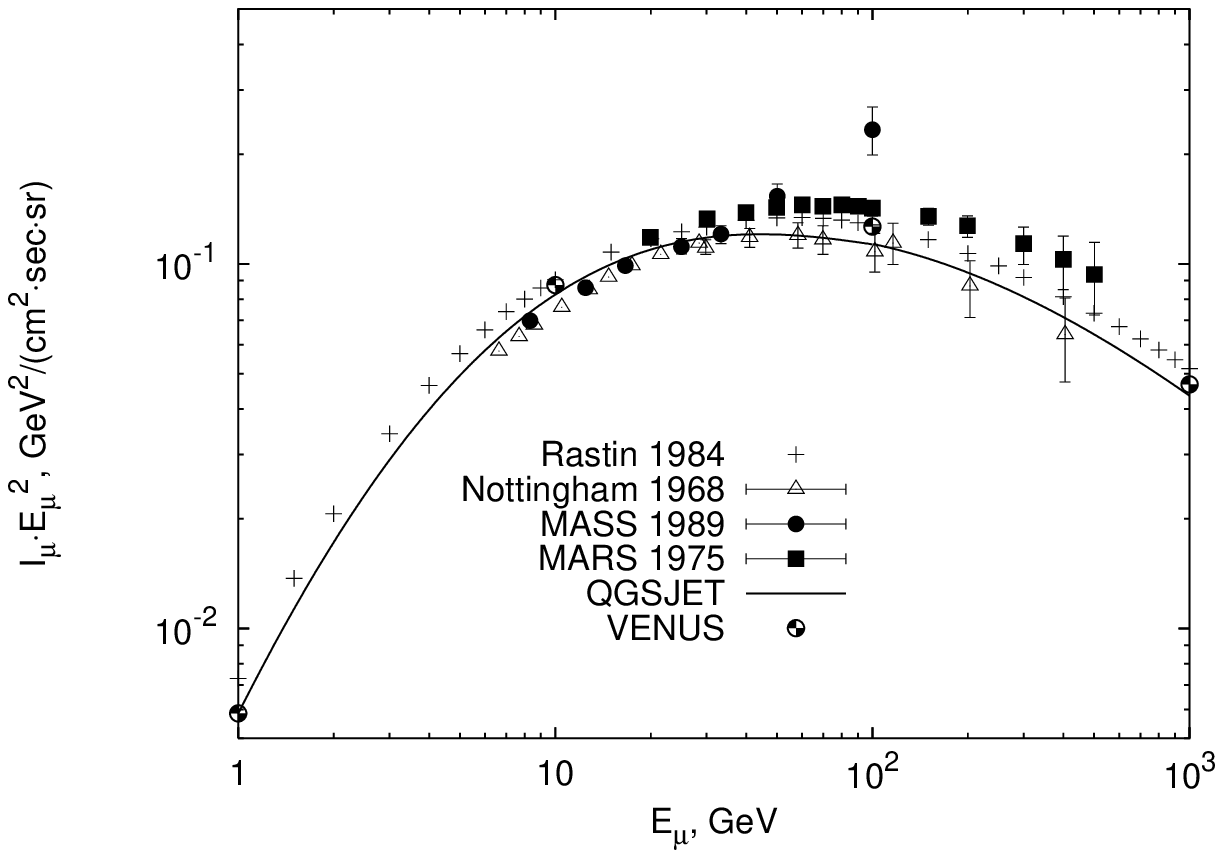}\hfill
\centering\includegraphics[height=0.4\textheight]{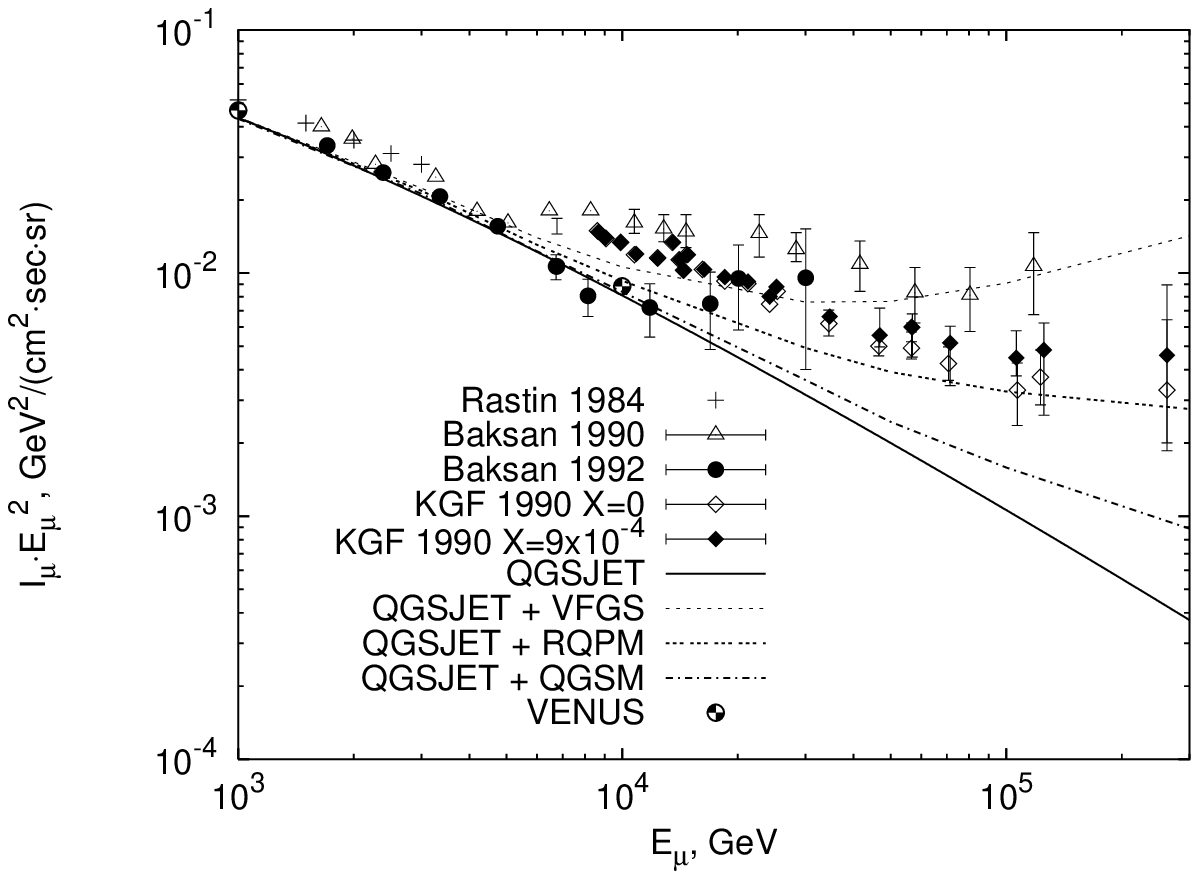}
\caption{Integral muon spectrum at sea level. QGSJET and
VENUS~---~present work calculations with the corresponding
interaction models for the primary nucleon spectrum, shown in
figure~\ref{nucleon_shift} with open circles. Other designations are
the same as in figure~\ref{diff}.} \label{aint}
\end{figure}

In conclusion, we would like to present an estimation of the changes
in nucleon flux, required to match the muon experimental data. In
figure~\ref{nucleon_shift} we show, along with initially applied
spectrum, its part for $E>2$~TeV 8\% shifted toward the higher
energies. From the magnet spectrometers region to the EC energies we
use an interpolation. So, muon spectra, given in
figures~\ref{adiff},\,\ref{aint}, are calculated with the use of the
combined nucleon flux, shown in figure~\ref{nucleon_shift} with open
circles. These spectra still present rather a lower estimate of the
muon data and may be approximated via the following formula (energy
is in GeV, $E_1=10$~GeV)
\begin{equation}
\label{seq_mu}
S_\mu(E)=AE^{(B+C\ln(E/E_1)+D\ln^2(E/E_1)+F\ln^3(E/E_1))},
\end{equation}
with parameters, given in table~3.

The used nucleon spectrum is just a model spectrum, since in
practice, correction to the PCR flux hardly may be constant and
should depend both on the energy and atomic mass of the projectile.
But it is easy to note, that required 8\% shift is significantly
smaller, than the quoted accuracy of the energy determination in the
EC experiments and, on the other hand, to some part it may be
attributed to the effects, discussed in this section.

\renewcommand{\tabcolsep}{3pt}
\begin{table}
\label{param} \caption{Parameters of approximation~(2) of muon
spectra at sea level.}
\begin{indented}\item[]
\begin{tabular}{@{}l*5{l}}
\multicolumn{6}{c}{}\\ \multicolumn{6}{c}{Integral spectrum
$(\mathrm{cm}^2\cdot\mathrm{sec}\cdot\mathrm{sr})^{-1}$}\\ \br
\multicolumn{1}{c}{$E$, GeV}&\multicolumn{1}{c}{$A$}
&\multicolumn{1}{c}{$B$} & \multicolumn{1}{c}{$C$}
&\multicolumn{1}{c}{$D$} & \multicolumn{1}{c}{$F$}\\ \mr
$\leq$100&$5.8392\cdot10^{-3}$&$-8.4964\cdot10^{-1}$&$-2.5331\cdot10^{-1}$&$6.7652\cdot10^{-3}$&$\phantom{-}3.3739\cdot10^{-3}$\\
$>100$&$6.5584\cdot10^{-3}$&$-9.1783\cdot10^{-1}$&$-2.3051\cdot10^{-1}$&$1.3452\cdot10^{-2}$&$-3.3109\cdot10^{-4}$\\
\br \multicolumn{6}{c}{}\\ \multicolumn{6}{c}{Differential spectrum
$(\mathrm{GeV}\cdot\mathrm{cm}^2\cdot\mathrm{sec}\cdot\mathrm{sr})^{-1}$}\\
\br \multicolumn{1}{c}{$E$, GeV}&\multicolumn{1}{c}{$A$}
&\multicolumn{1}{c}{$B$} & \multicolumn{1}{c}{$C$}
&\multicolumn{1}{c}{$D$} & \multicolumn{1}{c}{$F$}\\ \mr
$\leq$100&$1.7146\cdot10^{-3}$&\multicolumn{1}{c}{$-1.1620$}&$-4.1126\cdot10^{-1}$&$3.6839\cdot10^{-2}$&$-8.0472\cdot10^{-4}$\\
$>100$&$3.7984\cdot10^{-3}$&\multicolumn{1}{c}{$-1.5376$}&$-2.7868\cdot10^{-1}$&$1.6661\cdot10^{-2}$&$-4.1802\cdot10^{-4}$\\
\br
\end{tabular}
\end{indented}
\end{table}

\section{Conclusions}
For the first time a complete set of the most recent experimental
data on PCR nuclei spectra has been used in calculations of muon flux
at sea level in a wide energy range from 1 to $3\cdot10^5$~GeV. The
adequate description of cascade processes in the atmosphere has been
provided by the application of widely approved code CORSIKA along
with QGSJET and VENUS models. This makes our results free of the
possible errors, caused by a simplified approach to description of
hadronic interactions.

The main and a rather unexpected conclusion, which our investigation
has led to, is that the data on primaries conform to the muon data
only for rather low energies up to several tens of GeV. From these
energies a deficit of muons becomes evident, corresponding to a
deficit of primary nucleons with $E_{\mathrm{PCR}}\gtrsim1$~TeV. In
the earlier works of the other authors this problem was masked by the
uncertainties of the data on PCR fluxes and hadronic interactions,
existed those days. In a surprising way, refinement and gain of the
information have not approached us to the possibility of the more
accurate muon flux derivation, but, on the contrary, have revealed
that our today's, improved notions are in some part erroneous. This,
by the way, makes obsolete attempts to standardize the lepton fluxes
calculations in respect to the primary spectra, since, as our
consideration shows, in the large part the problem lies in
underestimation of PCR intensity. In our point of view, to remedy the
situation with the least consequences, the re-examination of the EC
experimental results with the application of different interaction
codes is required. As we have noted, in this connection VENUS
possesses with two positive features. In comparison with the other
models it predicts a larger number of muons at sea level, and in the
single hadronic interaction, less fraction of the energy is deposited
into electromagnetic component, but the influence of the latter
effect on the measurements results is not so simply to analyze, since
it may be compensated at the different stages of the EC data
processing. Another fact, that must be taken into account, is the
possible presence in PCR of sizable fraction of antiprotons. If the
emulsion chambers data would prove to be insensitive to the given
effects, only then the problem should be attributed to some
uncertainties or anomalies in the high-energy hadronic interactions.
In any case, our calculations have demonstrated, that today the fit
of the sea level muon data can not be derived directly from the PCR
data, and the reasons, leading to this puzzle, look unclear.

\ack

We thank the two anonymous referees for their constructive comments,
which helped us significantly improve the paper.

This work is supported in part by the UR grants No. 02.01.014 and No. 02.01.001.

\end{document}